\documentclass[a4paper,12pt]{article}
\pdfoutput=1
\usepackage{epsfig}
\usepackage{dsfont}
\usepackage{amssymb}
\usepackage{amsfonts}
\usepackage{amsmath}
\usepackage{blindtext}
\usepackage{hyperref}
\usepackage{esint}
\usepackage{euscript}
\usepackage{verbatim}
\usepackage{latexsym}
\usepackage{graphicx}
\usepackage{caption}
\usepackage{float}
\usepackage{subcaption}
\usepackage{bbm}
\usepackage{circuitikz}
\usepackage{listings}

\usepackage{tikz, pgfplots}
\usetikzlibrary{decorations}
\usepackage{caption}
\usetikzlibrary[positioning]
\usetikzlibrary{snakes}
\usepackage{amsmath,amssymb} 
\usepackage{booktabs}
\usepackage{hyperref}
\hypersetup{
	colorlinks=true,
	linkcolor=black,
	filecolor=red,      
	urlcolor=blue,
	citecolor=blue
} 

\newif\ifdtup

\catcode`\@=11

\@addtoreset{equation}{section}

\def\@normalsize{\@setsize\normalsize{15pt}\xiipt\@xiipt
\abovedisplayskip 14pt plus3pt minus3pt%
\belowdisplayskip \abovedisplayskip
\abovedisplayshortskip \z@ plus3pt%
\belowdisplayshortskip 7pt plus3.5pt minus0pt}

\def\small{\@setsize\small{13.6pt}\xipt\@xipt
\abovedisplayskip 13pt plus3pt minus3pt%
\belowdisplayskip \abovedisplayskip
\abovedisplayshortskip \z@ plus3pt%
\belowdisplayshortskip 7pt plus3.5pt minus0pt
\def\@listi{\parsep 4.5pt plus 2pt minus 1pt
     \itemsep \parsep
     \topsep 9pt plus 3pt minus 3pt}}

\relax

\catcode`@=12

\catcode`\@=11

\def\section{\@startsection{section}{1}{\z@}{3.5ex plus 1ex minus
   .2ex}{2.3ex plus .2ex}{\large\bf}}

\def\SymBoxes#1#2#3#4{\newdimen\un@t \un@t#3%
\raisebox{#1}{\rule{#2\un@t}{#4}\hskip-#2\un@t% lower horizontal
\@tempdimb\un@t \advance\@tempdimb by-#4\@tempcntb#2\relax%
\@whilenum{\@tempcntb>0}\do{%                         % #2 vertical lines
\rule{#4}{\un@t}\hskip\@tempdimb \advance\@tempcntb by\m@ne}%
\hskip-#2\un@t \rule[\un@t]{#2\un@t}{#4}%
\rule[\un@t]{#4}{#4}\hskip-#4%             % upper horizontal line
\rule{#4}{\un@t}}\hskip-#4}                % rightest vertical line
\begin{document}
%\begin{letter}{~}

%%%%%%Define some new commands and  macros
\newcommand{\beq}{\begin{equation}}
\newcommand{\eeq}{\end{equation}}
\newcommand{\bea}{\begin{eqnarray}}
\newcommand{\eea}{\end{eqnarray}}
\newcommand{\beas}{\begin{eqnarray*}}
\newcommand{\eeas}{\end{eqnarray*}}
\newcommand{\defi}{\stackrel{\rm def}{=}}
\newcommand{\non}{\nonumber}
\newcommand{\bquo}{\begin{quote}}
\newcommand{\enqu}{\end{quote}}
%%%%%%%%%%%%%%%%
\renewcommand{\(}{\begin{equation}}
\renewcommand{\)}{\end{equation}}
%%%%%%%%%%%%%%%%%%%%%%%%%%%%%%%%%% definitions
\def \eqn#1#2{\begin{equation}#2\label{#1}\end{equation}}

\def\e{\epsilon}
\def\IZ{{\mathbb Z}}
\def\IR{{\mathbb R}}
\def\IC{{\mathbb C}}
\def\IQ{{\mathbb Q}}
\def\de{\partial}
\def\Tr{ \hbox{\rm Tr}}
\def\H{ \hbox{\rm H}}
\def\HE{ \hbox{$\rm H^{even}$}}
\def\HO{ \hbox{$\rm H^{odd}$}}
\def\K{ \hbox{\rm K}}
\def\Im{ \hbox{\rm Im}}
\def\Ker{ \hbox{\rm Ker}}
\def\const{\hbox {\rm const.}}
\def\o{\over}
\def\im{\hbox{\rm Im}}
\def\re{\hbox{\rm Re}}
\def\bra{\langle}\def\ket{\rangle}
\def\Arg{\hbox {\rm Arg}}
\def\Re{\hbox {\rm Re}}
\def\Im{\hbox {\rm Im}}
\def\exo{\hbox {\rm exp}}
\def\diag{\hbox{\rm diag}}
\def\longvert{{\rule[-2mm]{0.1mm}{7mm}}\,}
\def\a{\alpha}
\def\dag{{}^{\dagger}}
\def\tq{{\widetilde q}}
\def\p{{}^{\prime}}
\def\W{W}
\def\N{{\cal N}}
\def\hsp{,\hspace{.7cm}}

\def\br{\nonumber}
\def\IZ{{\mathbb Z}}
\def\IR{{\mathbb R}}
\def\IC{{\mathbb C}}
\def\IQ{{\mathbb Q}}
\def\IP{{\mathbb P}}
\def \eqn#1#2{\begin{equation}#2\label{#1}\end{equation}}

\newcommand{\C}{\ensuremath{\mathbb C}}
\newcommand{\Z}{\ensuremath{\mathbb Z}}
\newcommand{\R}{\ensuremath{\mathbb R}}
\newcommand{\rp}{\ensuremath{\mathbb {RP}}}
\newcommand{\cp}{\ensuremath{\mathbb {CP}}}
\newcommand{\vac}{\ensuremath{|0\rangle}}
\newcommand{\vact}{\ensuremath{|00\rangle}                    }
\newcommand{\oc}{\ensuremath{\overline{c}}}
\newcommand{\psizero}{\psi_{0}}
\newcommand{\phizero}{\phi_{0}}
\newcommand{\hzero}{h_{0}}
\newcommand{\psiin}{\psi_{\rh}}
\newcommand{\phiin}{\phi_{\rh}}
\newcommand{\hin}{h_{\rh}}
\newcommand{\rh}{r_{h}}
\newcommand{\rb}{r_{b}}
\newcommand{\psibnd}{\psi_{0}^{b}}
\newcommand{\psibndp}{\psi_{1}^{b}}
\newcommand{\phibnd}{\phi_{0}^{b}}
\newcommand{\phibndp}{\phi_{1}^{b}}
\newcommand{\gbnd}{g_{0}^{b}}
\newcommand{\hbnd}{h_{0}^{b}}
\newcommand{\zh}{z_{h}}
\newcommand{\zb}{z_{b}}
\newcommand{\man}{\mathcal{M}}
\newcommand{\hbr}{\bar{h}}
\newcommand{\tbr}{\bar{t}}
\usetikzlibrary{arrows}
\newcommand{\midarrow}{\tikz \draw[-triangle 90] (0,0) -- +(.1,0);}
\begin{titlepage}
%\begin{flushright} CHEP XXXXX
%ULB-TH/09-10\\
%hep-th/yymmnnn\\ \end{flushright}
%\bigskip

\def\thefootnote{\fnsymbol{footnote}}

\begin{center}
{\large
{\bf Two dimensional de-Sitter and deformed CFTs}
}
\end{center}

%\bigskip
\begin{center}
\ Suchetan Das\footnote{\texttt{suchetan1993@gmail.com }} 
%\vspace{0.1in}

\end{center}

\renewcommand{\thefootnote}{\arabic{footnote}}

\begin{center}
%\vspace{0.2cm}
{School of Physical Sciences, Indian Association for the Cultivation of Science,\\
2A and 2B Raja S. C. Mullick Road, Jadavpur Kolkata-700032, India.}

 %{$^c$Department of Physics,\\ Ramakrishna Mission Vivekananda Educational and Research Institute,\\
%Belur Math, Howrah 711202, India.}

\end{center}
\vspace{-0.15in}
\noindent
\begin{center} {\bf Abstract}
\end{center}
We present an alternative dimensional reduction that yields an effective theory of dilatons in a two-dimensional de Sitter background. Specifically, by performing an S-wave reduction of higher-dimensional Einstein gravity, we obtain free massless dilatons in the Nariai static patch, and a dynamically evolving dilatons in the past Milne wedge. We then propose a (Nariai) static patch worldsheet formulation in terms of CFTs with SL(2,$\mathbb{R}$) deformed Hamiltonians on the cylinder. A key feature of this construction is that a stretched horizon in the (Nariai) static patch, equipped with an emergent UV boundary condition, acts as a gravitating observer. Using the similar reduction, we have also  obtained a Schwarzian action coupled to free massless dilatons in the near horizon near extremal limit of four dimensional charged AdS black holes. The worldsheet description for the same has been proposed and discussed in \cite{Das:2025cuq}. We also comment on how different notions of worldsheet time may themselves be \textit{emergent}.

\end{titlepage}

\setcounter{footnote}{0}
\tableofcontents

\section{Introduction}\label{sec1}
In a closed universe, `the problem of time' appeared to be the most challenging conceptual as well as technical aspect to formulate a quantum theory of gravity \cite{Kuchar:1991qf}-\cite{Carlip:2023daf}. In an open universe like AdS, where the ADM Hamiltonian acquires boundary term, one can define asymptotic observers and observables with an extrinsic notion of boundary time. In AdS/CFT, the UV completed boundary CFT thus has a well-defined intrinsic notion of time. Even though the notion of proper boundary time could be different for the action of large diffeomorphism in asymptotically AdS(specially at two and three dimensions), the boundary `external' observers can keep track of the history with their own ADM clock \cite{Wei:2025guh}. However, in closed universe like dS, general covariance of GR dramatically contradicts with the quantum mechanical notion of `fixed time' through Wheeler-Dewitt equation: $\hat{H}|\psi\rangle_{uni}=0$ \cite{DeWitt:1967yk}, \cite{Wheeler:1968iap}. Here the vanishing Hamiltonian constraints on physical state of the universe would dictates a `static universe' that directly contradicts with our observed experience. In other way, `the problem of time' indirectly suggests that time should be an emergent phenomenon in a quantum theory of gravity in a closed universe. 

One of the most compelling resolutions was proposed by Page and Wootters\cite{Page:1983uc}. According to it, for a closed quantum system subject to a Hamiltonian constraint (such as the Wheeler–DeWitt equation), an internal observer can experience dynamics via a `relational time,' arising from the system’s entanglement with a clock subsystem. More recently, this resolution has been recast as an addition of a gravitating observer in a closed universe with a goal towards defining a background independent formulation of quantum gravity\cite{Chandrasekaran:2022cip}-\cite{DeVuyst:2024khu}. Of course, by definition the observer should be \textit{emergent} in the full UV theory. To be more precise, in low energy limit, one may think of the total Hamiltonian of an observed system to be sum of the Hamiltonian of the subsystem $H_{S}$ and the Hamiltonian of the gravitating observer $H_{O}$ \cite{DeVuyst:2024khu}. In special cases\footnote{When we neglect the space-time fluctuation of gravitational subsystem in a sufficiently low energy.}, one may take $H_{S}$ (upto a multiplicative constant) as the modular Hamiltonian of the subsystem with respect to a cyclic, separating state vector $|\psi\rangle_{S}$. In this combined system, one can define $|\psi(t)\rangle_{S} = e^{-iH_{S}t}|\psi(0)\rangle_{S}$, where the projected $|\psi(0)\rangle_{S} \equiv \langle\phi(0)|\psi\rangle$ and $|\phi(t)\rangle_{O} = e^{-iH_{O}t}|\phi(0)\rangle_{O}$. In this way, imposing the constraint would tell us the observer measures \textit{modular time} defined by $H_{S}$ with respect to $|\psi\rangle_{S}$ on the subsystem $S$.  

Lately, there has been considerable work \cite{DeVuyst:2024khu}-\cite{Harlow:2025pvj}\footnote{This is not exhaustive, see the cited works for more.} exploring the idea of including the observer as part of the gravitational system in a closed universe following \cite{Witten:2023qsv}. The most natural setting to investigate this proposal is a closed de Sitter space-time, where a static observer’s accessible region is confined to a static patch bounded by cosmological (as well as black hole) horizons. Apart from a theoretical playground, dS universe has a more direct contact with observational cosmology and the fate of our living universe. The widely studied inflationary regime of cosmology has been approximately described by a metastable dS phase. On the other hand, the present observation of accelerated expansion also suggests that the universe will go through an asymptotic de-sitter phase at late time. If we imagine an astronomical observer of the current time, it will probably see a locally dS static patch in his own clock after sufficiently long time. Hence understanding the huge amount of `mysterious' Hawking-Gibbons entropy of static patch \cite{Gibbons:1977mu} and consequently providing a workable model of `static patch holography' would be a great deal of interest. %There are many versions of `static patch holography' can be found in the literature \cite{Gibbons:1977mu}. 
The most obvious notion may come from the stretched horizon holography, where in the absence of an asymptotic boundary, the vast amount of entropy could be located near the horizon\cite{Susskind:2021omt},\cite{Susskind:2021esx},\cite{Susskind:2023rxm}. More intuitively, from the perspective of a podal observer(located at the center of static patch), all the degrees of freedom may be captured by the timelike stretched horizon, where the stretched horizon itself plays the role of a gravitating observer\footnote{Note that, an interesting feature of the static patch cosmological horizon is it's observer dependence. Even though, different observer could see different horizon, the area is always be the same.}. Nevertheless, only a clean UV theory of static patch effective theory could lead us to the correct and proper understanding of static patch entropy and holography.

By definition, the UV theory should have a low energy limit that describes (semi)classical theory of gravity\footnote{This should be a minimal requirement for a consistent UV theory.}. In the closed universe, as we mentioned, the UV theory \textit{should not} have a \textit{definite notion of time}. Here, in a modest sense, our goal of this note is to find a sensible theory that only describes an observer in the lower dimensional dS static patch in low energy limit. To be more precise, we are interested to study two dimensional effective theory of dS in specific coordinates and it's possible UV description. With this motivation, we study S-wave dimensional reduction of asymptotic Schwarzchild dS$_{4}$ geometry in the near nariai near horizon limit to find an effective 2D description of dilatons in \textbf{section (\ref{sec2})}. In contrast to JT gravity, we provide an alternative reduction, in which the effective theory can be recast as free massless dilatons propagating in a static patch background enclosed by cosmological and blackhole horizon with same temperature. We quantize the dilatons using some \textit{UV dirichlet boundary condition} at the stretched horizon and find a finite entropy which we interpret as out-side entropy $S_{out}$ of observers living inside the static patch with a worldline along the stretched horizon. In \textbf{section (\ref{sec3})}, we consider S-wave reduction of pure dS$_{3}$ in 3D Einstein gravity. Unlike a free theory, this effective theory(from any higher dimensional reduction) can be written as dynamical dilatons inside a past Milne wedge\footnote{Here past Milne wedge refers to the past of $I^{+}$ or the upper wedge in the Penrose diagram as in fig (1.b) of \cite{Dey:2025osp}.}. We also observe the dilaton effective theory has an initial potential dominance and late time kinetic dominance, providing us a simple toy model for inflation in 2D. We relate it to a critical quench problem and then discuss a boundary non-preserving deformed CFT Hamiltonian on a strip in \textbf{section \ref{sec 2.1}}. The ultimate fate of the constructed past Milne vacuum is shown to be unstable at (finite but) late time, where the kinetic term dominates over potential term in \textbf{section (\ref{sec 2.2})}. Since Milne patch is analytically continued to static patch with thermal Euclidean time circle, we immediately constructs a sensible worldsheet description of (extended) static patch\footnote{Throughout the note, we refer `extended static patch' as the dimensionally reduced Nariai black hole enclosed by black hole and cosmological horizon at same temperature.} in terms of deformed CFT Hamiltonian on a ring in \textbf{section (\ref{sec4})}.  We further study it's modular quantization and connect the thermal entropy with entanglement entropy of static patch vacuum in \textbf{section (\ref{sec4.1}) and (\ref{sec4.2})}. The boundary condition on the stretched horizon is an emergent requirement of this quantization, as we also noticed in AdS \cite{Das:2024mlx},\cite{Das:2025cuq}. We also discuss our proposed alternative S-wave reduction in 4D magnetically charged RN AdS$_{4}$ blackhole in the near horizon near extremal limit and find free massless dilatons coupled to boundary Schwarzian as an effective description in \textbf{appendix (\ref{A})}. This provides a closure to our claim in a previous paper \cite{Das:2025cuq} on the UV completion of JT coupled to $c=\mathcal{O}(1)$ CFT as a worldsheet deformed CFT in a more concrete way.

However, in this modest approach, the questions regarding the observer dependent choice of UV time is still obscure, since our description of worldsheet CFT already knows the `time' of what observer will see in a static patch. We know in dS, different patches describe different notion of time and hence finding a global vacuum is inherently a difficult problem\footnote{See \cite{Witten:2023xze} and the discussion in \cite{Liu:2025krl} from algebra of observables point of view. However, our present understanding might be more aligned with \cite{Sen:2025bmj}, as well as \cite{Witten:2023xze}.}. We end our note with some comments on it in \textbf{section \ref{sec6}} along with possible future direction.

\section{Dimensional Reduction to 2D de-Sitter in near-Nariai limit}\label{sec2}

Here we review the dimensional reduction of pure Einstein gravity in 4D to two-dimensional JT gravity in the near-Nariai limit \cite{Nariai:1999iok},\cite{Ginsparg:1982rs} following \cite{Svesko:2022txo}. Consider the following metric ansatz of the class of spherically symmetric asymptotically 4D de Sitter metric:
\begin{align}\label{asymp met}
    ds^{2}_{4} = g_{\mu\nu}dx^{\mu}dx^{\nu} + r_{N}^{2}\phi(x) d\Omega^{2}_{2}
\end{align}
Here, $g_{\mu\nu}$ is an unspecified two-dimensional metric with $\mu,\nu=(0,1)$ and $\phi(x)$ is the dilaton field. The corresponding four-dimensional Einstein-Hilbert action is
\begin{align}\label{4D action}
    I_{4} = \frac{1}{16\pi G_{4}} \int_{M_{4}}d^{4}X \sqrt{-\hat{g}_{4}}(\hat{R}_{4}-2\Lambda); \; \Lambda = \frac{3}{L_{4}^{2}}
\end{align}
Here $X=(x_{\mu},\Omega_{2})$. $\hat{g}_{4}$ and $\hat{R}_{4}$ denote the determinant of the four-dimensional metric (\ref{asymp met}) and the corresponding Ricci scalar, respectively. $\hat{R}_{4}$ also reduces to the following form \cite{Svesko:2022txo} :
\begin{align}\label{R in 4D}
    \hat{R}_{4} = R_{2} + \frac{2}{r_{N}^{2}\phi} +\frac{(\nabla \phi)^{2}}{2\phi^{2}} - \frac{2}{\phi}\nabla^{2}\phi
\end{align}
Here, $R_{2}$ and $\nabla$ refer to a two-dimensional Ricci scalar and the covariant derivative with respect to the 2D metric $g_{\mu\nu}(x)$. Also, because of spherical reduction we have,
\begin{align}\label{4d to 2d}
    \int_{M_{4}}d^{4}X \sqrt{-\hat{g}_{4}} = r_{N}^{2}\Omega_{2}\int_{M_{2}} d^{2}x \sqrt{-g}\phi
\end{align}
By defining, $\frac{1}{G_{2}} \equiv \frac{r_{N}^{2}\Omega_{2}}{G_{4}}$ and by using (\ref{R in 4D}) and (\ref{4d to 2d}), the action (\ref{4D action}) reduces to the following 2D action
\begin{align}\label{effective 2d action 1}
    I_{4} = \frac{1}{16\pi G_{2}} \int_{M_{2}} d^{2}x \sqrt{-g} \left(\phi R_{2}-2\Lambda\phi+\frac{2}{r_{N}^{2}}+\frac{(\nabla\phi)^{2}}{2\phi}-2\nabla^{2}\phi\right)
\end{align}
In order to obtain a JT gravity action, the kinetic term must be dropped out. This can be achieved by a Weyl rescaling of the metric $g_{\mu\nu}\rightarrow \omega^{2}(x)\bar{g}_{\mu\nu}$. Under such Weyl rescaling, the metric determinant and $R_{2}$ transform as:
\begin{align}\label{weyl1}
  \sqrt{-g}=\omega^{2}  \sqrt{-\bar{g}} \; ; \;  R_{2} = \frac{\bar{R}_{2}}{\omega^{2}} - \frac{2\nabla^{2}\omega}{\omega^{3}}+\frac{2(\nabla\omega)^{2}}{\omega^{4}} .
\end{align}
Thus putting (\ref{weyl1}) back into (\ref{effective 2d action 1}), we end up with
\begin{align}\label{effective 2d 2}
   I_{4} = \frac{1}{16\pi G_{2}} \int_{\bar{M}_{2}} d^{2}x \sqrt{-\bar{g}}\left[\phi \bar{R}_{2} - \frac{2\phi\nabla^{2}\omega}{\omega} +\frac{2\phi(\nabla\omega)^{2}}{\omega^{2}}-2\Lambda\phi\omega^{2}+\frac{2\omega^{2}}{r_{N}^{2}}+ \frac{(\nabla\phi)^{2}}{2\phi}-2\nabla^{2}\phi\right]
\end{align}
Note that, here the covariant derivative $\nabla$ is taken with respect to $\bar{g}_{\mu\nu}$. Now we can write
\begin{align}
    &\frac{2\phi(\nabla\omega)^{2}}{\omega^{2}} = \nabla^{\mu}(\frac{2\phi}{\omega}\nabla_{\mu}\omega)-\frac{2}{\omega}(\nabla^{\mu}\phi)(\nabla_{\mu}\omega)-\frac{2\phi}{\omega}\nabla^{2}\omega-2\phi\omega\nabla^{\mu}(\frac{1}{\omega^{2}})\nabla_{\mu}\omega \nonumber \\
  & \implies  \frac{2\phi(\nabla\omega)^{2}}{\omega^{2}}  = \frac{2\phi}{\omega}\nabla^{2}\omega + \frac{2}{\omega}(\nabla^{\mu}\phi)(\nabla_{\mu}\omega) -\nabla^{\mu}(\frac{2\phi}{\omega}\nabla_{\mu}\omega)
\end{align}
Inserting the above identity in (\ref{effective 2d 2}), we have
\begin{align}\label{scaled action}
  I_{4} = \frac{1}{16\pi G_{2}} \int_{\bar{M}_{2}} d^{2}x \sqrt{-\bar{g}}\left[\phi \bar{R}_{2} + \frac{2}{\omega}(\nabla^{\mu}\phi)(\nabla_{\mu}\omega)-2\Lambda\phi\omega^{2}+\frac{2\omega^{2}}{r_{N}^{2}}+ \frac{(\nabla\phi)^{2}}{2\phi}-2\nabla^{2}\phi - 2\nabla^{\mu}(\frac{\phi}{\omega}\nabla_{\mu}\omega) \right]   
\end{align}
If we now choose $\omega=\frac{\alpha}{\phi}$, we can eliminate the total derivative term as well as the kinetic term, since
\begin{align}
    \frac{2}{\omega}(\nabla^{\mu}\phi)(\nabla_{\mu}\omega)|_{\omega=\frac{\alpha}{\phi}} = -\frac{(\nabla\phi)^{2}}{2\phi}; \; 2\nabla^{\mu}(\frac{\phi}{\omega}\nabla_{\mu}\omega)|_{\omega=\frac{\alpha}{\phi}} = -2\nabla^{2}\phi
\end{align}
Hence the choice simplifies the effective 2D action as
\begin{align}
    I_{4} = \frac{1}{16\pi G_{2}} \int_{\bar{M}_{2}} d^{2}x \sqrt{-\bar{g}}\left[\phi \bar{R}_{2} -\frac{2\Lambda\alpha^{2}}{\phi}+\frac{2\alpha^{2}}{r_{N}^{2}\phi^{2}}\right]
\end{align}
Note that, this choice of scaling factor is slightly different than \cite{Svesko:2022txo}, in which the total derivative terms in the effective action are getting canceled with the boundary Gibbons-Hawking term. However, here we are not adding a GHY boundary term in a closed universe setting. Hence the potential term will also be different from theirs. However the end result is similar. Using $\Lambda=\frac{3}{L_{4}^{2}}$, the Nairai black hole radius $r_{N}=\frac{L_{4}}{\sqrt{3}}$ and choosing $\alpha=\frac{1}{\sqrt{3}}$ we finally obtain
\begin{align}
    I_{4} = \frac{1}{16\pi G_{2}} \int_{\bar{M}_{2}} d^{2}x \sqrt{-\bar{g}}\left[\phi \bar{R}_{2} +U(\phi)\right]; \; \text{where} \; U(\phi) = \frac{2}{L_{4}^{2}}\left(-\frac{1}{\phi}+\frac{1}{\phi^{2}}\right)
\end{align}
Once again, 
\begin{align}
    U(\phi=1) = 0, \; \frac{dU}{d\phi}|_{\phi=1} = -\frac{2}{L^{2}_{4}}
\end{align}
Thus by expanding $\phi=\phi_{0}+\Phi$ with $\phi_{0}=1$, the leading order term of $I_{4}$ provides the effective JT gravity action as
\begin{align}\label{jt from 4d}
    I_{4} \approx I_{JT} = \frac{1}{16\pi G_{2}} \int_{\bar{M}_{2}} d^{2}x \sqrt{-\bar{g}}\phi_{0}\bar{R}_{2} + \frac{1}{16\pi G_{2}} \int_{\bar{M}_{2}} d^{2}x \sqrt{-\bar{g}}\Phi\left[\bar{R}_{2}-\frac{2}{L_{4}^{2}}\right]
\end{align}
One can readily recognize the 1st term as the topological Gauss-Bonnet term without boundary, whereas the second term belongs to the standard JT gravity with positive cosmological constant. One can connect it with the standard reduction of JT gravity from 4D Schwarzchild de-Sitter(SdS) black holes in near Nariai limit. The metric of 4D SdS black hole is the following \cite{Svesko:2022txo}:
\begin{align}\label{sds}
    ds^{4}=-f(r)dt^{2}+\frac{dr^{2}}{f(r)}+r^{2}d\Omega^{2}_{2}; \; \text{where} \; f(r) = 1-\frac{r^{2}}{L_{4}^{2}}-\frac{16\pi G_{4}M}{2\Omega_{2}r} = \frac{(r-r_{h})(r_{c}-r)(r+r_{h}+r_{c})}{rL_{4}^{2}}.
\end{align}
$M$ is the black hole mass. Here $f(r)$ has two positive root: $f(r_{h})=f(r_{c})=0$, where $r_{h}$ and $r_{c}$ are black hole and cosmological horizon respectively. Also $r_{h}<r<r_{c}$. In terms of these horizon radius, we can express \cite{Svesko:2022txo}
\begin{align}
    L_{4}^{2} = \frac{r_{c}^{3}-r_{h}^{3}}{r_{c}-r_{h}}, \; \frac{16\pi G_{4}M}{2\Omega_{2}}= \frac{r_{h}r_{c}^{3}-r_{h}^{3}r_{c}}{r_{c}^{3}-r_{h}^{3}}.
\end{align}
The Nariai limit is defined as $r_{h}=r_{c}\equiv r_{N}$. Correspondingly,$r_{N} = \frac{L_{4}}{\sqrt{3}}$. However in this coordinate system, taking Nariai limit is ill-defined as $f(r) \rightarrow 0$ when $r_{c}\rightarrow r_{h}$, since $r_{h}<r<r_{c}$. As it is reviewed in \cite{Svesko:2022txo}, one can use the following change of coordinate to zoom in the near horizon limit where such limit can exist.
\begin{align}
 \tau = \frac{\beta\epsilon t}{2\hat{L}_{4}}, \; \rho= \frac{2\hat{L}_{4}}{\beta}\left(\frac{r-r_{h}}{\epsilon}-\frac{\beta}{2}\right), \; \beta=\frac{r_{c}-r_{h}}{\epsilon}, \; \hat{L}_{4}=\frac{L_{4}}{\sqrt{3}}=r_{N}.   
\end{align}
Taking $\epsilon \rightarrow 0$ limit, the metric (\ref{sds}) reduces to dS$_{2} \times$ S$^{2}$:
\begin{align}\label{Nariai BH}
    ds^{2} = -\left(1-\frac{\rho^{2}}{r_{N}^{2}}\right)d\tau^{2} + \left(1-\frac{\rho^{2}}{r_{N}^{2}}\right)^{-1}d\rho^{2}+r_{N}^{2}d\Omega_{2}^{2}
\end{align}
Here $\rho\in[-r_{N},r_{N}]$. This coincide with (\ref{asymp met}) where $\phi=1$ and $g_{\mu\nu}$ corresponds to 2D de-Sitter in static patch. 

We can then study the effective JT action (\ref{jt from 4d}) and it's quantization using standard gravitational path integral procedure \cite{Maldacena:2019cbz},\cite{Cotler:2019nbi}. The usual approach is to treat the JT gravity itself as an effective theory of 2D gravity and then proceed to the rigorously well-defined gravitational path integral with dilaton boundary condition as an UV input. For instance, one can simply compute the Hartle-Hawking wave function by doing explicit path integral of the form
    \begin{align}
        \psi_{+}[h] \propto \int_{\bar{g}_{\partial M}=h,\Phi_{\partial M}=\phi_{b}} D\bar{g}_{\mu\nu} D\Phi e^{-I_{JT}}
    \end{align}
    This is relevant for studying model of 2D quantum gravity independent of it's origin in higher dimension. However, this type of path integral is only meaningful over Hartle-Hawking type geometries, where one can specify certain boundary condition on metric as well as dilaton \cite{Cotler:2019nbi} at conformal infinity. %Even though, the static patch is analytically continued version of Milne patch which has future or past boundary, a direct path integral interpretation for such static patch is still missing. Also one need to include `boundary' GHY term at conformal boundary, which generates the Schwarzian action from certain boundary condition on dilaton and metric near the timelike infinity.
    However when considering the static patch, one can recover the total entropy as follows \cite{Svesko:2022txo}-\cite{Galante:2023uyf}:
    \begin{align}
        S = \frac{2\phi_{0}}{4G_{2}}+\frac{\Phi_{bh}}{4G_{2}}+\frac{\Phi_{c}}{4G_{2}}
    \end{align}
 Since $\Phi_{bh}=-\Phi_{c}$, the net entropy becomes the entropy of Nariai black hole, i.e. $S=S_{N}=\frac{2\phi_{0}}{4G_{2}}$. Without adding such boundary GHY term, the path integral becomes trivial as it fixes $\bar{R}_{2}=\frac{2}{L^{2}_{4}}$ and $\bar{g}_{\mu\nu}$ is fixed to be locally dS$_{2}$ metric. Hence the sole contribution to the entropy coming only from $\phi_{0}$ or Nariai term and the near Nariai correction does not contribute unlike JT in AdS$_{2}$.
 
\subsection{Alternative dual reduction in the near-Nariai limit}\label{sec 2.1}

 Another approach is to take care of the dimensional reduction and quantizing the effective theory governed by 4D gravitational path integral. Since we use a Weyl rescaling $g_{\mu\nu} = \omega^{2}\hat{g}_{\mu\nu}$ and then choose $\omega$ to be some function of $\Phi$, the effective theory has only dynamical variable $\Phi$ once we fix the background $\hat{g}_{\mu\nu}$ to be that of (\ref{Nariai BH}) in the near Nariai limit \cite{Easson:2025ekn}. By exploiting conformal flatness of any two-dimensional metric, we can always make this choice which further restricts the form of $R_{2}$ in terms of dilatons. Since the dilatons are arbitrary at this moment, we can always do such fixing. Thus we can fix $\hat{R}_{2} = \frac{2}{\hat{L}_{4}^{2}}$. Finally we are left with only
 \begin{align}
     &Z = e^{-I_{0}}\int D\Phi e^{-\frac{1}{4\pi G_{2}L^{2}_{4}}\int d\rho d\tau_{E}\Phi(\rho,\tau_{E})} =A e^{-I_{0}}, \nonumber \\ \text{where} \; &I_{0} = \frac{\phi_{0}}{16\pi G_{2}}\int^{2\pi \hat{L}_{4}}_{0}d\tau_{E} \int^{\hat{L}_{4}}_{-\hat{L}_{4}}d\rho \frac{2}{\hat{L}^{2}_{4}}=\frac{\phi_{0}}{2G_{2}}, \; \text{and} \; A=\int D\Phi e^{-\frac{1}{4\pi G_{2}L^{2}_{4}}\int d\rho d\tau_{E}\Phi(\rho,\tau_{E})}
 \end{align}
Hence, the entropy is $S=S_{N}+\log(A)$. This suggests, the total entropy may have some correction which is not observed from previous way. However, this is not a useful way to study dynamics of the effective theory. This motivates us to consider an Weyl-equivalent or alternative  dimensional reduction which may capture another useful and tractable description of near horizon physics in closed universe.  

To derive JT action, we have used a particular scaling form $\omega$. In stead, we could choose any other $\omega=f(\phi)$, which may cancel the $\frac{(\nabla\phi)^{2}}{2\phi}$ term in the action (\ref{scaled action}). For instance, choosing $\omega= c\phi^{-\frac{1}{4}}e^{\beta\phi}$ in (\ref{scaled action}) yields
\begin{align}
    I'_{4} = \frac{1}{16\pi G_{2}} \int_{\bar{M}_{2}}d^{2}x \sqrt{-\bar{g}}\left[ \phi\bar{R}_{2}+2\beta(\nabla\phi)^{2}+2\Lambda c^{2}e^{2\beta\phi}\left(\frac{1}{\sqrt{\phi}}-\sqrt{\phi}\right)-\nabla^{2}\phi -2\beta\nabla^{\mu}(\phi\nabla_{\mu}\phi)\right]
\end{align}
Here $\beta$ is a dimensionful constant. This is a \textit{Liouville-type} action with an effective potential
\begin{align}
    U'(\phi) = 2\Lambda c^{2}e^{2\beta\phi}\left(\frac{1}{\sqrt{\phi}}-\sqrt{\phi}\right), \; \text{such that:}\; U'(\phi=1)=0, \frac{dU'}{d\phi}|_{\phi=1} = -2\Lambda c^{2}e^{2\beta}
\end{align}
If we further choose $\beta=\frac{1}{4}$ and $c=e^{-\frac{1}{4}}$, after expanding $\phi=\phi_{0}+\Phi$ with $\phi_{0}=1$, we end up with
\begin{align}
    I'_{4} =\frac{\phi_{0}}{16\pi G_{2}} \int_{\bar{M}_{2}}d^{2}x \sqrt{-\bar{g}}\bar{R}_{2}+ \frac{1}{16\pi G_{2}} \int_{\bar{M}_{2}}d^{2}x \sqrt{-\bar{g}}\left[\frac{1}{2}(\nabla\Phi)^{2}+ \Phi\left(\bar{R}_{2}-\frac{2}{\hat{L}^{2}_{4}}\right)-\frac{3}{2}\nabla^{2}\Phi -\frac{1}{2}\nabla^{\mu}(\Phi\nabla_{\mu}\Phi)\right]
\end{align}
Using the same logic which we explained earlier, we could always fix the two-dimensional $\bar{g}_{\mu\nu}$ to be the static patch metric with $\bar{R}_{2}=\frac{2}{\hat{L}^{2}_{4}}$. Note that by doing this one would find a relation between $R_{2}, \frac{2}{\hat{L}^{2}}$ and $\phi$. Hence, by varying $\phi$, we could in principle get all possible $g_{\mu\nu}$ to be consistent with (\ref{asymp met}). Thus, we end up with
\begin{align}\label{dual JT1}
    I'_{4} = I_{0}+\frac{1}{16\pi G_{2}} \int d\rho d\tau \left[\frac{1}{2}(\nabla\Phi)^{2}-\frac{3}{2}\nabla^{2}\Phi -\frac{1}{2}\nabla^{\mu}(\Phi\nabla_{\mu}\Phi)\right]
\end{align}
This is a free scalar field action upto two total derivative terms. Those terms can be vanishing if we use the following boundary condition at the two horizons and the usual periodic boundary condition at $(\tau_{E}=0,2\pi\hat{L}_{4})$ in the Euclidean setting:
\begin{align}
    \Phi(\rho,\tau_{E})|_{\rho=\pm \hat{L}_{4}} = 0, \; \nabla_{\tau}\Phi(\rho,\tau_{E})|_{\tau_{E}=0}=\nabla_{\tau}\Phi(\rho,\tau_{E})|_{\tau=2\pi\hat{L}_{4}}, \; \Phi(\rho,0)=\Phi(\rho,2\pi\hat{L}_{4})
\end{align}
The periodic boundary condition is natural from the periodic nature of thermal Euclidean time circle. However, the Dirichlet boundary condition on $\Phi$ is non-standard and is not a direct requirement of the classical action. At this classical level, it seems to be ad-hoc to use such boundary condition on dilatons at the horizons(or more correctly on the stretched horizons). However, as we will argue later, one must impose such boundary condition in order to quantize a \textit{deformed CFT on a ring} which further describes CFT on extended static patch background. Hence we may think of this as an \textit{emergent boundary condition originated from certain UV theory having a worldsheet description in terms of deformed CFT on a ring}. In this alternative dimensional reduction, the near Nariai dynamics can be recast as  free massless scalar field in dS$_{2}$ static patch background with Dirichlet boundary condition at the horizons. %We note that, the spectrum of quantized scalar field contains certain primaries, the \textit{vertex operator} of the form $e^{ib\Phi}$. The Neumann boundary condition does not affect such operators at the horizon \footnote{On the other hand, if we further use Dirichlet boundary condition $\Phi|_{r=r_{N}}=0$, vertex operators on the horizon would be identity. However, the present problem does not require to impose Dirichlet boundary condition at horizons.}. We will see later, those operators will play important role in the entropy counting. 
In this reduction, the total entropy is
\begin{align}
    S=\frac{\phi_{0}}{2G_{2}}+S_{out}^{\Phi},
\end{align}
where $S_{out}^{\Phi}$ is the entropy of the fields $\Phi$. We may think of this as a generalized entropy of the extended static patch\footnote{In contrast to pure de-Sitter static patch, this contains both black hole and cosmological horizon as reduced from the Nariai limit of SdS black holes in 4D.} in dS$_{2}$, which is well-defined in the algebraic formulation of gravity by including observers in the static patch \cite{Chandrasekaran:2022cip}. The role of the observer should be evident while computing the explicit entropy of dilatons as we do in next section. %Here, in stead of an external observer action at the centre of static patch($\rho=0$), we have fields $\Phi$ all over the static patch. We will argue the thermal entropy of the scalar fields can be identified with the entanglement entropy of a spatial slice covering the full static patch bounded by cosmological and black hole horizon in section \ref{sec5}. This entanglement entropy is measured by an observer located on the pole who has causal access to the full static patch.
\subsection{Partition function of massless free scalar fields in a static patch background with stretched horizon}
%Explicitly compute free massless scalar field partition function in 2D static patch using brick wall regulator--should be easy in 2D. Relate to 2501.17912(even though it was done for massive field with no IR divergence problem), comment on 2511.01400 (no phase problem?)

%*************************************************
It would be useful to perform a coordinate change to map the static patch into conformally flat spacetime coordinate. Using Tortoise-like coordinate $\theta$ defines as
\begin{align}
    d\theta = \frac{d\rho}{1-\frac{\rho^{2}}{r_{N}^{2}}} \implies \left[ \frac{\theta}{r_{N}}=\frac{1}{2}\ln\frac{1+\frac{\rho}{r_{N}}}{1-\frac{\rho}{r_{N}}}=\tanh^{-1}\frac{\rho}{r_{N}}\right],
\end{align}
the metric of the static patch becomes
\begin{align}\label{static in conformal}
    ds^{2}=\frac{-d\tau^{2}+d\theta^{2}}{\cosh^{2}(\frac{\theta}{r_{N}})}
\end{align}
Assuming a separation of variable by exploiting time translational symmetry $\Phi(\theta,\tau)=\psi_{\omega}(\theta)e^{-i\omega\tau}$, the solution of massless Klein-Gordon equation $\nabla^{2}\Phi(\theta,\tau)=0$ is
\begin{align}
    \psi_{\omega}(\theta)= Ae^{i\omega\theta}+Be^{-i\omega\theta}
\end{align}
To impose Dirichlet boundary condition at the horizons, we use brick wall regularization pioneered by t'hooft \cite{tHooft:1984kcu}. We consider streteched horizons located at $\rho=r_{N}-\epsilon$ and $\rho=-r_{N}+\epsilon$. Correspondingly in conformal frame, \begin{align}
    \frac{\theta}{r_{N}}|_{\rho=r_{N}-\epsilon}=\frac{1}{2}\ln(\frac{2r_{N}}{\epsilon})\equiv \Lambda_{c}, \; \frac{\theta}{r_{N}}|_{\rho=-r_{N}+\epsilon}=\frac{1}{2}\ln(\frac{\epsilon}{2r_{N}})\equiv -\Lambda_{c}
\end{align}
Hence imposing Dirichlet condition at stretched horizons, we end up with $A=\pm B$. Choosing $A=-B$, the solution becomes $\Phi=A\sin(\omega\theta)e^{-i\omega\tau}$. The Dirichlet boundary condition further implies \cite{Soni:2023fke}
\begin{align}\label{normal modes}
    \sin\omega r_{N}\Lambda_{c} = 0 \rightarrow \; \omega_{n}=\frac{2n\pi}{r_{N}\ln(\frac{2r_{N}}{\epsilon})}, n\in Z
\end{align}
This form exactly coincides with the \textit{emergent} discrete Virasoro modes of UV deformed CFT as we will describe in section \ref{sec5}. Choosing $A=-B$ would give $(2n+1)\pi/2\Lambda_{c}$ which does not coincide with the UV theory requirement. Note that, one could also give same form of normal mode by imposing Neumann boundary condition. Also one can check the boundary term $[\partial_{\theta}\Phi(\theta=\Lambda_{c})-\partial_{\theta}\Phi(\theta=-\Lambda_{c})]$ exactly vanishes \footnote{The other term will also be vanishing by Dirichlet condition itself.}.

To fix the normalization $A$, one can further use Klein-Gordon inner product as in \cite{Burman:2023kko} which provides an orthonormal basis of mode functions. This is also needed to construct the Hilbert space for the fields. One can easily check that $A=\frac{1}{\sqrt{2\omega_{n}}}$. To compute the partition function, this normalization is crucial. Since we know the exact normal modes, we can perform now an exact computation of partition function as in \cite{Burman:2023kko},\cite{Krishnan:2023jqn} \footnote{See also \cite{Banerjee:2024ivh}.}after quantizing the fields. We can write the operator $\Phi$ in terms of normalized modes $\psi_{\omega_{n}}$ and the creation and annihilation operators as:
\begin{align}
    \Phi(\theta,\tau) = \sum_{n=0}^{\infty}\frac{1}{\sqrt{2\omega_{n}}}\left[a_{n}\psi_{\omega_{n}}(\theta)e^{-i\omega_{n}\tau}+a_{n}^{\dagger}\psi_{\omega_{n}}(\theta)e^{i\omega_{n}\tau}\right]
\end{align}
Here $a_{n}$ and $a_{n}^{\dagger}$ satisfies canonical commutation relation and also defines the `Boulware vacuum' $|0\rangle_{B}$ by \cite{Mukohyama:1998rf},\cite{Burman:2023kko}
\begin{align}
    a_{n}|0\rangle_{B} = 0
\end{align}
Hence the normal ordered Hamiltonian is also given by
\begin{align}
    :H: = \sum_{n}\omega_{n}a_{n}^{\dagger}a_{n}
\end{align}
The Hilbert space can now be obtained by considering all eigenstates of the Hamiltonians which os of the known form:
\begin{align}
    |n\rangle = \prod_{n}\frac{1}{\sqrt{N_{n}}}(a^{\dagger})^{N_{n}}|0\rangle_{B}, \; \text{such that} \; :H:|n\rangle = \sum_{n}(\omega_{n}N_{n})|n\rangle
\end{align}
Here $N_{n}$ is the occupation number at level $N$. Thus the thermal partition function of the dilatons $\Phi$s in thermal equilibrium with the static patch of inverse temperature $\beta_{N}$\footnote{To be precise the temperature of the static patch is given by $T_{N}=\frac{1}{2\pi r_{N}}$ \cite{Svesko:2022txo}.} is given by
\begin{align}
    Z_{N}(\beta_{N}) = Tr(e^{-\beta_{N}:H:}) = \prod_{n}\sum_{N_{n}}e^{-\beta_{N}\omega_{n}N_{n}} = \prod_{n}\frac{1}{1-e^{-\beta_{N}\omega_{n}}}
\end{align}
Thus taking $\log$ we get
\begin{align}
    \log(Z_{N}) = -\sum_{n}\log(1-e^{-\beta_{N}\omega_{n}})
\end{align}
In $\epsilon\rightarrow 0$ limit, the discrete normal mode spectrum becomes continuous $\omega_{n}\rightarrow \omega$ and we can convert the sum into integral as follows:
\begin{align}
    \log(Z_{N}) = -\frac{r_{N}\log(\frac{2r_{N}}{\epsilon})}{2\pi}\int^{\infty}_{0}d\omega\log(1-e^{-\beta_{N}\omega}) = \frac{\pi r_{N} \log(\frac{2r_{N}}{\epsilon})}{12\beta_{N}}=\frac{\log(\frac{2r_{N}}{\epsilon})}{24}
\end{align}
Correspondingly the thermal entropy becomes
\begin{align}\label{out entropy}
    S^{\Phi}_{out} = \frac{1}{24}\log(\frac{2r_{N}}{\epsilon})
\end{align}
In section (\ref{sec5}) we will explicitly compute and will try to match this entropy using the quantization of our proposed UV free bosonic deformed CFT on a ring. We note that, the Hamiltonian of dilatons with stretched horizon boundary condition defined by some \textit{UV theory}, should be identified with the total Hamiltonian of the gravitating observer plus dilatons and the observer clock is set by the `time' generated by static patch Hamiltonian due to Hamiltonian constraint.
%(Naive discu\ssion)

%Hence in this alternative reduction in the near-Nariai limit, we obtain a different theory of dilatons. However, if we compute the partition function by integrating over all possible values of dilatons (satisfying Neumann boundary condition), we should obtain the same entropy in the semiclassical limit as we obtained in JT gravity description. Since by construction, we have only used a different scaling as a function of $\phi$, integrating over all $\phi$ should give the same partition function in these two apriori different descriptions. One of the main goals of this note is to reproduce the static patch entropy using this dual description of (\ref{dual JT1}). This will be studied in section \ref{}. Hence the crux claim of this section is that \textit{JT gravity has a dual interpretation in terms of free scalar fields in the static patch background with Neumann boundary condition. These two theories are dual in the sense that}
%\begin{align}
    %\lim_{G_{2}\rightarrow 0} \sum_{\text{topologies}}\int D g_{\mu\nu}D\Phi e^{-I_{JT}} \approx e^{-S_{N}}=\lim_{G_{2} \rightarrow 0}\int D\Phi e^{-I'_{4}}
%\end{align}
%\textit{In the semiclassical limit, the genus zero contribution only dominates in the sum over topologies. We also expect the same duality holds beyond the semiclassical limit.}
\section{Dimensional reduction to 2D de-Sitter from Einstein gravity in pure dS$_{D\geq 3}$}\label{sec3}
%problem in half reduction to get static patch, reduction to Milne patch and near (late time)boundary free CFT, connecting to quench problem
Unlike higher dimension $(D\geq 4)$, an effective theory in 2D de-Sitter can also be obtained from Einstein gravity with positive cosmological constant in three dimension \cite{Kames-King:2021etp},\cite{Svesko:2022txo}. For such reduction, one need to consider pure de-Sitter metrics in 3D. To obtain the effective theory, we start with the following ansatz of the 3D metrics \cite{Svesko:2022txo}:
\begin{align}
    ds^{2} = g_{\mu\nu}dx^{\mu}dx^{\nu}+ L_{3}^{2}\phi^{2}(x)d\Omega^{2}_{1}
\end{align}
The Einstein-Hilbert action is
\begin{align}
    I_{3} = \frac{1}{16\pi G_{3}}\int_{M} d^{3}X \sqrt{-\hat{g}_{3}}[\hat{R}_{4}-2\Lambda_{3}], \; \Lambda_{3}=\frac{1}{L_{3}^{2}}
\end{align}
Since $\hat{R}_{3}=R_{2}-\frac{2}{\phi}\nabla^{2}\phi$, we end up with the usual JT gravity up to a total derivative term
\begin{align}\label{jt from 3d}
    I_{3} = \frac{1}{16\pi G_{2}}\int_{g}d^{2}x \sqrt{-g}[\phi R-2\Lambda\phi-2\nabla^{2}\phi], \; \frac{1}{G_{2}}\equiv\frac{2\pi L_{3}}{G_{3}}
\end{align}
If one start with the solution of Einstein equation in 3D as pure dS$_{3}$ in static patch coordinate:
\begin{align}
    ds^{2} = -\left(1-\frac{r^{2}}{L_{3}^{2}}\right)dt^{2}+\frac{dr^{2}}{\left(1-\frac{r^{2}}{L^{2}_{3}}\right)}+r^{2}d\theta^{2},
\end{align}
one should end up with the JT gravity(upto total derivative term) in dS$_{2}$ static patch:
\begin{align}
    ds^{2} = -\left(1-\frac{r^{2}}{L_{3}^{2}}\right)dt^{2}+\frac{dr^{2}}{\left(1-\frac{r^{2}}{L^{2}_{3}}\right)}
\end{align}
Here $\phi=\frac{r}{L_{3}}$. It is noted that \cite{Svesko:2022txo}, for such pure de-Sitter reduction, $\phi$ must be positive which ensure the positivity of effective Newton's constant and hence $0<r<L_{3}$. Thus, only half of the full de-Sitter, denoted as the static patch accessible to the observers located on two poles, is described by this metric. This geometry is described by a different Penrose diagram from that of the static patch reduced from higher dimensional near Nariai solution, as noted in \cite{Svesko:2022txo}. 

One now observe that in the absence of a GHY term, the sole contribution to the gravitational path integral for JT-dilaton gravity should come from the total derivative or `boundary term'. In terms of the angular coordinate $r=L_{3}\cos\theta$, the static patch covers $-\frac{\pi}{2}<\theta<\frac{\pi}{2}$. From the Penrose diagram these two are the south and north pole\cite{Svesko:2022txo}. Hence choosing a suitable `boundary' condition on these poles should determine the static patch entropy due to the presence of cosmological horizon. However, poles are not special points and hence can not be treated as `boundary' of the space-time. Thus choosing some fixed values of dilatons(or it's derivatives) at those points are strictly ill-defined. At most, one can impose finiteness of the solutions at those points coming from self-adjointness of strum-Liouville type differential operator\footnote{See \cite{Stone:1} (sec 8.2.1) for a related discussion in the context of solving Laplace equation in spherical polar coordinates.}. We will now perform an analogous Weyl rescaling of 2D metrics as we did in the previous section, to rewrite the effective action in a useful form motivated from the previous section.

Using a Weyl scaling $g_{\mu\nu} \rightarrow \omega^{2}(x)\bar{g}_{\mu\nu}$, (\ref{jt from 3d}) reduces to the following:
\begin{align}
    I_{3} = \frac{1}{16\pi G_{2}}\int d^{2}x \sqrt{-\bar{g}_{2}}\left[ \phi\bar{R}_{2}+\frac{2}{\omega}(\nabla^{\mu}\phi)(\nabla_{\mu}\omega)-2\Lambda\phi\omega^{2}-2\nabla^{2}\phi-2\nabla^{\mu}\left(\frac{\phi}{\omega}\nabla_{\mu}\omega\right)\right]
\end{align}
However choosing a similar $\omega$ as in the previous section, we can not recover free field action. In stead choosing $\omega=ce^{\frac{\phi}{4}}$, we end up with the following form:
\begin{align}
    I_{3} = \frac{1}{16\pi G_{2}}\int d^{2}x \sqrt{-\bar{g}_{2}}\left[ \frac{(\nabla\phi)^{2}}{2}+\bar{R}_{2}\phi-2c\Lambda\phi e^{\frac{\phi}{2}}-2\nabla^{2}\phi-\frac{1}{2}\nabla^{\mu}\left(\phi\nabla_{\mu}\phi\right)\right]
\end{align}
This type of Liouville-type action (upto total derivative terms) was first observed by Solodukhin in \cite{Solodukhin:1998tc} in the context of emergent near horizon conformal symmetry. The potential of the form $\phi e^{\frac{\phi}{2}}$ makes the theory non-conformal which can be thought of as a deformed Liouville action. By studying the equation of motion one can check non vanishing trace of stress tensor\footnote{One should note that, the Liouville CFT on a curved manifold can acquire non vanishing constant trace of stress tensor $T=Tr(T_{\mu\nu})$, which is proportional to $\bar{R}_{2}$. However, for this deformed Liouville, apart from the trace anomaly term, $T$ also accompanies with terms involving $\phi$, which makes it strictly non-conformal.} as in \cite{Solodukhin:1998tc}. Again we can choose $\bar{g}_{\mu\nu}$ to be locally dS$_{2}$ without loss of any generality. We consider $\bar{g}$ to be the Milne patch, related to static patch by analytic continuation \cite{Kames-King:2021etp}, having the following form of metric \cite{Cadoni:2002kz}:
\begin{align}\label{Milne to static Lor}
    ds^{2}_{Milne} &=-dT^{2}+\sinh^{2}(\frac{T}{L_{3}})d\theta^{2},  \; (\text{analytic continuation:}\;r=L_{3}\cosh(\frac{T}{L}),t=\theta)\\
    &=\frac{1}{\sinh^{2}(\frac{s}{L_{3}})}(-ds^{2}+d\theta^{2}) , \;  \; \text{using}\;\cosh(\frac{T}{L_{3}}) = \coth(\frac{s}{L_{3}})\\
    & = -\frac{d\hat{t}^{2}}{\hat{t}^{2}-L_{3}^{2}}+(\hat{t}^{2}-L_{3}^{2})d\theta^{2},\; \text{using} \; \coth(\frac{s}{L_{3}})=\frac{\hat{t}}{L_{3}}
\end{align}
Here the range of time in these coordinate systems are 
\begin{align}
  -\infty<T<0, \;  0<s<\infty, \; L_{3}<\hat{t}<\infty
\end{align}
The past conformal boundary corresponds to $T \rightarrow -\infty, s\rightarrow 0,\hat{t}\rightarrow \infty$. Hence in this patch, the late time $s \rightarrow \infty$ should correspond to $\hat{t} \rightarrow L_{3}$. This is the past horizon in static patch. In this late time limit for the past Milne patch, the metric becomes
\begin{align}\label{late time milne}
    ds^{2}_{Milne}|_{s\rightarrow\infty} \approx e^{-\frac{2s}{L_{3}}}(-ds^{2}+d\theta^{2})
\end{align}
%This is the same metric as late time limit of FLRW metric in two dimension. 
Now the effective action in past Milne patch becomes:
\begin{align}
    I_{3} = \frac{1}{16\pi G_{2}}\int ds d\theta \left[ -\frac{1}{2}(\partial_{s}\phi)^{2}+\frac{1}{2}(\partial_{\theta}\phi)^{2}+\frac{2\phi}{L_{3}^{2}\sinh^{2}(\frac{s}{L})}(1-ce^{\frac{\phi}{2}})+\text{Total Derivative}\right]
\end{align}
At past conformal boundary $s\rightarrow 0$, the potential term should dominate over the kinetic term. However, in the late time limit $s \rightarrow \infty$ of past Milne wedge with a time band cut-off $\Lambda_{IR}<s<\infty$, the action becomes a free massless scalar field theory(upto total derivative terms) in the approximate geometry (\ref{late time milne}). Hence the classical solution of $\phi$ can be written as $\phi=\int dk T_{k}(s)e^{ik\theta}$, where $T_{k}(s)=a_{1}e^{iks}+a_{2}e^{-iks}$. The exact solution is determined by imposing the boundary condition at $(s=\Lambda_{IR},\infty)$. The boundary condition should be an UV input and from the classical action we do not have any freedom to choose a meaningful one. Since specifying boundary condition on future slice would affect the initial boundary condition, it can make a potential violation of causality\cite{Anninos:2011jp}. The correct way is to specify initial condition and evolve it with the correct evolutionary operator\footnote{That necessarily would be an UV input.} capturing the full dynamics; which could further tell us the correct `boundary condition' on the approximate time band.%The physical requirement will be the non-normalizable part should be treated as a source of the total derivative or boundary term at the horizon of Milne wedge . %In this way, the problem of future boundary condition on dilatons might be replaced by suitable boundary theory sourced by non-normalizable bulk dilaton modes. 

We end this section by making two comments:
\begin{itemize}
    \item The problem of imposing unphysical `boundary' condition on poles in the static patch can be controlled by going into the past Milne patch and by using late time effective conformal theory as we explained above. However, it is still unclear about the correct physical boundary condition one should impose at the boundary of the time band purely from this classical action. We note that, the similar analysis can be carried out for higher dimensional reduction of pure de-Sitter geometries. Following the similar way\footnote{The details are not present in this present note to avoid unnecessary cluttering.}, one could obtain an effective theory of the past Milne patch with a different potential term. Interestingly,\textit{ one again obtains an universal feature showing the early time dominance of potential term(of different form for 2d reduction from different dimensional Einstein gravity) and the late time dominance of free bosonic CFT.}  

    \item The effective dynamical theory in the past Milne patch of being non conformal to conformal one, can be regarded as some kind of time dependent RG flow. More intuitively, one may map this into certain quench problem, where initial gapped theory is evolved by a gapless CFT Hamiltonian after certain Lorentzian time ($\Lambda_{IR}$). This will motivate us to study certain deformed CFT Hamiltonian in the next section, which breaks time translation symmetry upon acting on an initial vacuum state of a gapped theory at the past infinity and describes time dependent dynamics of the Milne patch.

\end{itemize}

\section{Deformed CFT, Milne patch and critical quench}\label{sec4}
Effectively, to capture the effective Milne patch dynamics(full) obtained in the past section, one need to find a time dependent evolution operator that is best described in a \textit{smooth critical quench} set-up\cite{Das:2014jna}-\cite{Das:2017sgp}. Given the Lagrangian of the effective theory, it seems analytically less tractable in such a set-up. However, in a crude sense, we can still approximate the dynamics as an instantaneous critical quench protocol \cite{Sengupta2004}, which enables us to study post quench dynamics inside the approximate time band $s:(\Lambda_{IR},\infty)$. In \cite{Calabrese:2006rx}, Cardy and Calabrese pioneered an analytically tractable critical quench set-up, where the initial state is explicitly written down as $|\psi\rangle_{in}=e^{-\tau_{0}H_{CFT}}|B\rangle$, which evolves to a late time steady state under post quench CFT evolution \cite{Calabrese:2016xau}. Here $\tau_{0}$ encodes the mass-gap parameter of the initial gapped Hamiltonian. $|B\rangle$ refers to the conformal boundary state. The justification of choosing such state comes from the RG sense where pre-quench and post-quench Hamiltonian is related by small irrelevant deformation under which any boundary condition flows to a conformal boundary condition. A relatively \textit{nicer} argument was presented in \cite{Das:2021qsd}\footnote{We thank Bobby Ezhuthachan for explaining this.}, where the Euclidean time strip picture is more naturally appeared from analytic continuation of Euclidean correlators to Lorentzian one via $i\epsilon$ prescription. To mention it briefly, consider the correlation function
\begin{align}
    \langle\psi|e^{iHt_{1}+\epsilon_{1}H}\mathcal{O}(x_{1})e^{-iHt_{12}-\epsilon_{12}H}\mathcal{O}(x_{2})e^{-iHt_{23}-\epsilon_{23}H}\dots \mathcal{O}(x_{n})e^{-iHt_{n}-\epsilon_{n}H}|\psi\rangle
\end{align}
Here we used the analytic continuation $\tau \rightarrow it+\epsilon$ and the notation $t_{ij}=t_{i}-t_{j}$. According to the $i\epsilon$ prescription, this correlation function is analytic when $(0>\epsilon_{1}>\epsilon_{2}>\dots >\epsilon_{n}>0)$, which does not make any sense. However this would not be a problem when $|\psi\rangle$ is an eigenstate of $H$. The problem occurs only when the initial state is not an eigenstate of the Hamiltonian. The effective way to get rid of this problem, is to formally introduce a cut-off $\tau_{0}$ to regulate high energy modes ($>\frac{1}{\tau_{0}}$). Hence the regulated state should be $e^{-\tau_{0}H}|\psi\rangle$, for which the analytic continuation has the following order: $\tau_{0}>\epsilon_{1}>\epsilon_{2}>\dots>\epsilon_{n}>-\tau_{0} $. In this Euclidean picture, the correlation function is computed in a strip of Euclidean time width $2\tau_{0}$. The boundary condition at the edges of this strip is chosen to be conformal one following the CC ansatz. 

To map this set-up with the effective Milne patch quench, we may think $\tau_{0}$ to be certain scale dictated by the non conformal initial theory. Here cleverly, the problem of imposing boundary condition in the effective Lorentzian time width is replaced by imposing conformal boundary condition at the edges of the strip in Euclidean time. This no longer violates any causal requirement of the original Lorentzian problem. Nevertheless, the Hamiltonian of the CFT is still needed to be the same of that of a free massless bosons in the approximate geometry (\ref{late time milne}) of late time band in past Milne wedge. This will motivate us to find a deformed CFT Hamiltonian, that captures the dynamics of Milne wedge as well as mimics the set-up of the critical sudden quench.

\subsection{Boundary non-preserving deformed CFT Hamiltonian on a strip and Milne patch dynamics}
Since PSL(2,$\mathbb{R}$) is the isometry group of dS$_{2}$, the metrics are conformally related to Euclidean AdS$_{2}$ metrics. For instance, to find Milne wedge in conformally flat coordinates, one may find a conformal mapping from EAdS$_{2}$ Poincare to dS$_{2}$ in Milne wedge. Consider the map $z=\mu\coth{\mu\omega}, \bar{z}=\mu\coth{\mu \bar{\omega}}$ and the EAdS$_{2}$ Poincare metric becomes
\begin{align}\label{Poincare to Milne}
    ds^{2}=4L^{2}_{AdS}\frac{dzd\bar{z}}{(z-\bar{z})^{2}} = 4L^{2}_{AdS} \left(\frac{\partial z}{\partial \omega}\right)\left(\frac{\partial \bar{z}}{\partial \bar{\omega}}\right)\frac{d\omega d\bar{\omega}}{(z(\omega)-\bar{z}(\bar{\omega}))^{2}} \rightarrow 4L^{2}_{AdS}\mu^{2} \frac{ds_{E}^{2}+d\theta^{2}}{\sin^{2}2\mu s_{E}}
\end{align}
Here $z=t+ix$,$\bar{z}=t-ix$ and $\omega=\theta+is_{E}$, $\bar{\omega}=\theta-is_{E}$. After analytically continuing $s_{E}\rightarrow is$ and $L_{AdS}\rightarrow iL$\footnote{Here $L$ is the de-Sitter radius.} with setting $2\mu L=1$, one could end up with the standard Milne wedge metric described in the previous section:
\begin{align}
    ds^{2} = 4L^{2}\mu^{2}\frac{-ds^{2}+d\theta^{2}}{\sinh^{2}{2\mu s}} = \frac{-ds^{2}+d\theta^{2}}{\sinh^{2}{\frac{s}{L}}}
\end{align}
In \cite{Das:2025cuq}, we have constructed SL(2,$\mathbb{R}$) deformed Hamiltonian on a strip, which describes one-sided AdS$_{2}$ black hole. Similarly, the natural question is to find a deformed Hamiltonian which manifests such role reversal of Euclidean spacetime $t\rightarrow is_{E}, x \rightarrow i\theta$ upon going from AdS$_{2}$ to dS$_{2}$. Let us consider the following deformed Hamiltonian on a strip $(w,\bar{w})$ of width $\pi$\footnote{Here $w=\sigma+iu,\;\bar{w}=\sigma-iu$. The boundary of the strips are located at $u=0,\pi$}:
\begin{align}
    H_{o}=\frac{\gamma}{\pi}\int^{\pi}_{0}du\; \sin(u) T_{00}(u), \; \gamma\in \mathbb{R}
\end{align}
After mapping to Upper Half Plane (UHP) using
the map ($w \rightarrow z=e^{w}=t+ix$), we end up with the Hamiltonian on the UHP:
\begin{align}\label{defomred Ham}
    H=i\gamma(L_{1}-L_{-1})-i\gamma(\bar{L}_{1}-\bar{L}_{-1})
\end{align}
Here $L_{n},\bar{L}_{n}$ are defined as
\begin{align}\label{Fourier mode}
    &L_{n}=\frac{c}{24}\delta_{n,0} + \frac{1}{2}\int^{\pi}_{0}\frac{du}{\pi} e^{i n u}T(u) = \frac{c}{24}\delta_{n,0} + \frac{1}{2\pi i} \int_{C}dz z^{n+1}T(z)  \\
    &\bar{L}_{n}=\frac{c}{24}\delta_{n,0} + \frac{1}{2}\int^{\pi}_{0}\frac{du}{\pi} e^{-i n u}\bar{T}(u) = \frac{c}{24}\delta_{n,0} + \frac{1}{2\pi i} \int_{\bar{C}}d\bar{z} \bar{z}^{n+1}\bar{T}(\bar{z}).
\end{align}
Here, the contours $C$ and $\bar{C}$ refer to a half circle around the origin of UHP. For details, we refer \cite{Das:2025cuq}. 

The Hamiltonian (\ref{defomred Ham}) is Hermitian since $L_{n}^{\dagger}=L_{-n}$. However, the same Hamiltonian \textit{does not preserve the symmetry of the UHP and the conformal boundary condition}. One can simply check it by taking the Euclidean space-time generators $(l_{n}=z^{n+1}\partial_{z},\bar{l}_{n}=\bar{z}^{n+1}\partial_{\bar{z}}$ corresponding to the Virasoro modes $(L_{n},\bar{L}_{n})$ and act on the real boundary of the UHP as:
\begin{align}
   i\gamma(l_{1}-l_{-1}-\bar{l}_{1}+\bar{l}_{-1})(z-\bar{z})|_{z=\bar{z}} \neq 0 
\end{align}
Since $H$ is not a symmetry of the UHP, it should act non-trivially on the SL(2,$\mathbb{R}$) invariant vacuum. %Note that, the Hamiltonian (\ref{defomred Ham}) can be interpreted as \textit{modular momentum}\cite{Mintchev:2025yso} which generates translation between fixed point and boundary and being orthogonal to modular flow\footnote{In \cite{Czech:2017zfq}, \cite{Das:2019iit}, it is denoted as $P_{D}$.}

As in \cite{Das:2022pez}, to find the proper conformal frame of the Euclidean time($s_{E}$) evolution generated by $H_{o}$, we 
need to find the curves $(z(s),\bar{z}(s))$ solving the curve equation:
\begin{align}
    i\gamma(l_{1}-l_{-1}-\bar{l}_{1}+\bar{l}_{-1}) =\frac{\partial z}{\partial s}\frac{\partial}{\partial z}+\frac{\partial \bar{z}}{\partial s}\frac{\partial}{\partial \bar{z}}
\end{align}
This involves solving
\begin{align}
    \int \frac{dz}{i\gamma(z^{2}-1)}=\int ds_{E}
\end{align}
Similar integral equation exists for $\bar{z}$ with $i\rightarrow -i$. Solving this, we get the following curves:
\begin{align}
z=-\coth\gamma(\theta+is_{E})\equiv-\coth{\gamma\omega}, \; \bar{z} = -\coth{\gamma(\theta-is_{E})}\equiv-\coth{\gamma\bar{\omega}}.
\end{align}
Here $\theta$ is an integration constant of the curve equation. The boundary of the UHP $z=\bar{z}$ maps to the following:
\begin{align}
    z=\bar{z} \implies s_{E}=(0,\frac{\pi}{\gamma})
\end{align}
Hence the effective geometric description of $(\omega,\bar{\omega})$ coordinates is an Euclidean time-strip of width $\frac{\pi}{\gamma}$ and $\theta:(-\infty,\infty)$. To find the constant $s_{E}$ and constant $\theta$ curve in the UHP, we consider the following:
\begin{align}
    \frac{z-1}{z+1} = e^{2\gamma(\theta+is_{E})}, \; \frac{\bar{z}-1}{\bar{z}+1}=e^{2\gamma(\theta-is_{E})}
\end{align}
From this, to study the constant $\theta$ curve or the evolution of $s_{E}$, we would consider the following:
\begin{align}
    e^{4\gamma\theta} = \frac{(z-1)(\bar{z}-1)}{(z+1)(\bar{z}+1)}
\end{align}
This can be simplified to the following equation:
\begin{align}
    (t+\coth{2\gamma\theta})^{2}+x^{2} = (\operatorname{csch}2\gamma\theta)^{2}
\end{align}
Since $x\geq 0 $, the constant $\theta$ curves generates semicircles of radius $\operatorname{csch}2\gamma\theta$ centered around $(t=-\coth{2\gamma\theta},\;x=0)$. Thus the range of $s_{E}:(0,\pi/\gamma)$. At $\theta\rightarrow\pm \infty$, the curves shrinks to the point $t=\pm 1$. Consequently, to find the $\theta$ curves of constant $s_{E}$ curves, we consider the following:
\begin{align}
    e^{4i\gamma s_{E}} = \frac{(z-1)(\bar{z}+1)}{(z+1)(\bar{z}-1)}
\end{align}
After some algebraic manipulation, we end up with the following curve equation:
\begin{align}
    t^{2}+(x-\cot{2\gamma s_{E}})^{2}=(\csc 2\gamma s_{E})^{2}
\end{align}
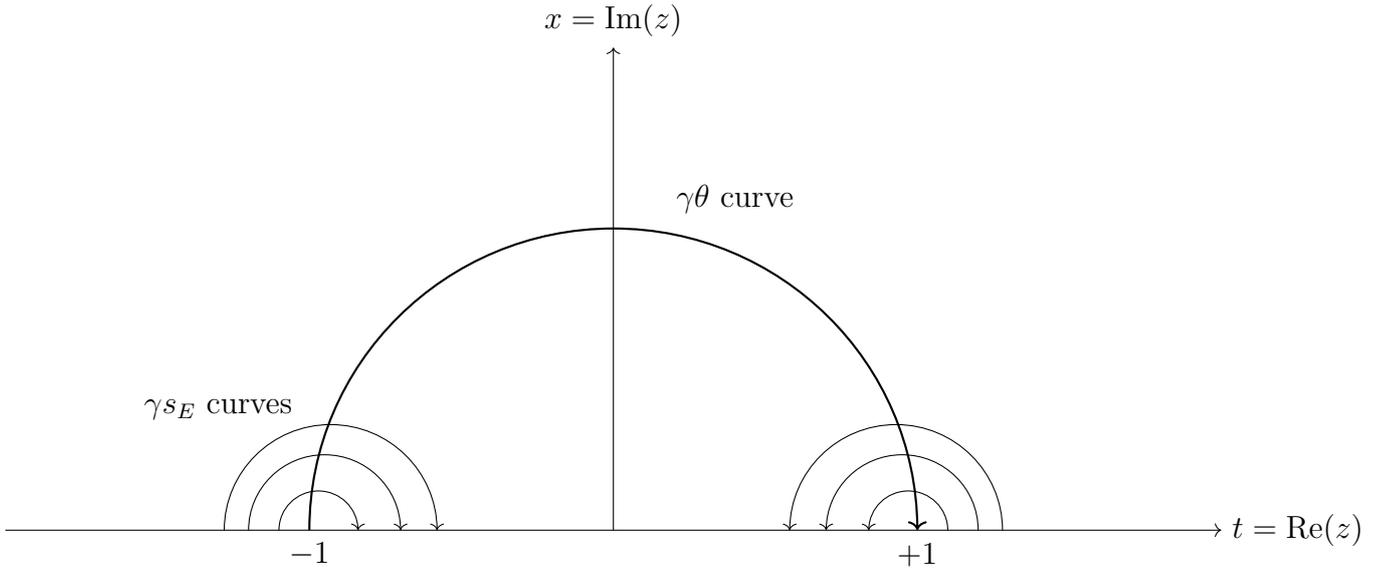
\begin{figure}
    \centering

\begin{tikzpicture}[scale=4]

% Axes
\draw[->] (-2,0) -- (2,0) node[right] {$t=\operatorname{Re}(z)$};
\draw[->] (0,0) -- (0,1.6) node[above] {$x=\operatorname{Im}(z)$};

% Labels -1 and +1
\node at (-1,-0.08) {$-1$};
\node at (1,-0.08) {$+1$};
\node at (0.4,1.1) {$\gamma\theta$ curve};
\node at (-1.3,0.4) {$\gamma s_{E}$ curves};
% Large semicircle from -1 to +1 (center 0)
\draw[thick,->] (-1,0) arc (180:0:1);
%\draw[thick,->] (-1,0) arc (180:0:1);
% Left side semicircles centered at -1
\draw[->] (-1.1,0) arc (180:0:0.13);
\draw[->] (-1.2,0) arc (180:0:0.25);
\draw[->] (-1.28,0) arc (180:0:0.35); % big one reaches to 0

% Right side semicircles centered at +1
\draw[->] (1.1,0) arc (0:180:0.13);
\draw[->] (1.2,0) arc (0:180:0.25);
\draw[->] (1.28,0) arc (0:180:0.35); % big one reaches to 0

\end{tikzpicture}
\caption{Curves of $\gamma\theta$ and $\gamma s_{E}$ on the UHP($t,x$)}
    \label{strip}
\end{figure}

One can see that $t=\pm 1$ and $x=0$ is the trivial solution of the curve for all $s_{E}$. Hence $(t=\pm 1,\; x=0)$ are the \textit{fixed point} of the Hamiltonian flow. Hence constant $s_{E}$ or the $\theta$ curves are semicircles end on the fixed points. This translates the fixed points to be $\theta=\pm \infty$. The curves are shown in the fig(\ref{strip}). 

One can also obtain the fixed points by solving $z(s_{E})=z(0)$. Here $z(s)=\coth{\gamma}(\theta+is_{E})$ and $z(0)=\coth{\gamma\theta}$. From this we get
\begin{align}
    z(s) = \frac{z(0)\cos{\gamma s_{E}}+i\sin{\gamma s_{E}}}{iz(0)\sin{\gamma s_{E}}+\cos{\gamma s_{E}}} \equiv \frac{a z(0)+b}{c z(0)+d} \; \text{with} \; ad-bc=1
\end{align}
Hence, the flow is mapped to a Mobius transformation with imaginary $b,c$. The solution of $z(s)=z(0)$ is $z=\pm 1$. This is the fixed point we got earlier which corresponds to $\theta= \pm \infty$. We note that this type of Hamiltonian flow has been studied in the context of Floquet CFTs \cite{Das:2023xaw},\cite{Banerjee:2024zqb}. In particular, \cite{Banerjee:2024zqb} studied a Floquet drive for which the dynamical \textit{heating phase} has similar structure of Mobius transformation\footnote{In the same paper \cite{Banerjee:2024zqb}, non-trivial dynamics in entanglement asymmetry has been found for a similar driven CFT on a strip. The initial global symmetry is broken by the boundary and the boundary non-preserving Floquet drive would never restore the symmetry. \textit{This result and related observation provides one of the major motivations behind this present note.}}. 

Following the observation of \cite{Das:2025cuq}, we should consider the metric of UHP as the AdS$_{2}$ Poincare. Hence the effective metric under such flow becomes the Euclidean Milne wedge in de-Sitter as we described in (\ref{Poincare to Milne}). Upon analytic continuation to $s_{E}\rightarrow is$, we will get Lorentzian Milne patch with $s:(0,\infty)$. To study the time dependence, one may consider one-point function of CFT scalar primary of dimension $h$ in $(\omega,\bar{\omega})$ plane, by mapping it to UHP.
\begin{align}
    \langle \mathcal{O}(\omega,\bar{\omega})\rangle = \left(\frac{\partial z}{\partial \omega}\right)^{h}\left(\frac{\partial \bar{z}}{\partial \bar{\omega}}\right)^{h}\langle \mathcal{O}(z,\bar{z})\rangle_{UHP} = \left(\frac{a_{\mathcal{O}}\gamma}{i\sin{2\gamma s_{E}}}\right)^{2h}
\end{align}
Here $a_{\mathcal{O}}$ is an undetermined constant. By doing $s_{E}\rightarrow is$, we get 
\begin{align}
    \langle \mathcal{O}(s,\theta)\rangle = \left(\frac{\gamma}{\sinh{2\gamma s}}\right)^{2h}
\end{align}
Note that the late (real)time behavior coincides with the one point function under quantum critical quench \cite{Calabrese:2016xau}, which decays exponentially at late time $s>>\frac{1}{2\gamma}$. 

However, unlike the CC ansatz, here we took slightly different initial state. Our initial state has the following form:
\begin{align}
    |\psi\rangle_{in} = e^{-H_{o}}|0\rangle_{S}
\end{align}
Here $H_{o}$ is the deformed Hamiltonian (\ref{defomred Ham}) on strip and $|0\rangle_{S}$ is the SL(2,$\mathbb{R}$) invariant vacuum of the strip. In \cite{Das:2025cuq}, we identify the vacuum $|0\rangle_{S}$ with the vacuum of Poincare AdS$_{2}$. In other way, we may think of this $|\psi\rangle_{in}$ to be constructed by \textit{downlifting\footnote{ The non-conformal vacuum is downlifted with respect to the AdS$_{2}$ vacuum which has vanishing dilaton stress tensor in the Poincare patch.} the AdS$_{2}$ vacuum with the deformed Hamiltonian which captures the dynamics of dilatons in the past Milne wedge}.

\subsection{Emergence of steady state at late time}\label{sec 2.2}
We have now seen the deformed CFT on the strip has profound similarity with global quench set-up followed by CC ansatz. To be more precise, the current set-up should correspond to an \textit{inhomogeneous critical quench}\cite{Sotiriadis:2008ila},\cite{Das:2021qsd}. Here $1/\gamma$ plays the role of the cut-off $\tau_{0}$. However a subtle difference arises in the energy momentum tensor. While, in the CC state, the $\langle T\rangle_{CC}= \frac{\pi c}{24(2\tau_{0})^{2}}>0$, in our prescribed \textit{non-conformal} Milne vacuum:
\begin{align}
    {}_{in}\langle \psi| T(\omega) |\psi\rangle_{in} = -\frac{c\gamma^{2}}{12}<0
\end{align}
Since we are interested in the physical $\gamma \rightarrow \infty$ limit, the stress energy becomes negatively divergent. This would further indicates the state $|\psi\rangle_{in}$ is \textit{unstable}. To make further support on this, we will now compute two point primary correlation function on a fixed time slice $s$, following \cite{Calabrese:2016xau},\cite{Aramthottil:2021cov}:
\begin{align}
    &{}_{in}\langle \psi|\mathcal{O}(s_{E},0)\mathcal{O}(s_{E},\theta)|\psi\rangle_{in} \nonumber \\
    &= \left(\frac{\partial z_{1}}{\partial \omega_{1}}\right)^{h}\left(\frac{\partial \bar{z}_{1}}{\partial \bar{\omega}_{1}}\right)^{h}\left(\frac{\partial z_{2}}{\partial \omega_{2}}\right)^{h}\left(\frac{\partial \bar{z}_{2}}{\partial \bar{\omega}_{2}}\right)^{h}\langle \mathcal{O}(z_{1},\bar{z}_{1})\mathcal{O}(z_{2},\bar{z}_{2})\rangle_{UHP}
\end{align}
Here $z_{1,2}=\coth{\gamma\omega_{1,2}}$, $\bar{z}_{1,2}=\coth{\gamma\bar{\omega}_{1,2}}$ and
\begin{align}\label{points}
    \omega_{1}=is_{E},\bar{\omega}_{1}=-is_{E}; \; \omega_{2}=\theta+is_{E},\bar{\omega}_{2}=\theta-is_{E}.
\end{align}
The UHP two point correlator has the following structure
\begin{align}
  \langle \mathcal{O}(z_{1},\bar{z}_{1})\mathcal{O}(z_{2},\bar{z}_{2})\rangle_{UHP} = \left(\frac{z_{12}z_{2\bar{1}}}{z_{12}z_{\bar{1}\bar{2}}z_{1\bar{1}}z_{2\bar{2}}}\right)^{2h}F(\eta)  
\end{align}
Here the notation $z_{ij}=z_{i}-z_{j}$ and $F(\eta)$ is an undetermined function of cross ratio $\eta$. For free scalar CFT, one can exactly compute this two point function. However, here we consider general CFT to find universal late time behavior. Using (\ref{points}), we find
\begin{align}
    \eta = \frac{\sin^{2}{(2\gamma s_{E})}}{\sinh{\gamma(\theta-2is_{E})}\sinh{\gamma(\theta+2is_{E})}} 
\end{align}
After analytic continuation $s_{E}\rightarrow is$, we get
\begin{align}
    \eta = \frac{1-\cosh{4\gamma s}}{\cosh{2\gamma\theta}-\cosh{4\gamma s}}
\end{align}
We now consider the late time limit as well as $s,\theta>>\frac{1}{4\gamma}$\footnote{We are interested only in this limit, since the effective dilaton theory becomes approximately conformal only at late time.}. We can notice two following different behavior of $\eta$:
\begin{align}
   & \eta \sim 1, \; \text{when} \; s>\frac{\theta}{2} \\
   & \eta \sim e^{4\gamma(s-\frac{\theta}{2})} \sim 0 \; \text{when}\; s<\frac{\theta}{2}
\end{align}
We now compute
\begin{align}
 \left(\frac{z_{12}z_{2\bar{1}}}{z_{12}z_{\bar{1}\bar{2}}z_{1\bar{1}}z_{2\bar{2}}}\right)^{2h} = \gamma^{4h}\left[\frac{\cosh{2\gamma\theta}-\cosh{4\gamma s}}{\sinh^{2}{\gamma\theta}\sinh^{2}{2\gamma s}}\right]^{2h}   
\end{align}
Again considering $s,\theta>>\frac{1}{4\gamma}$, we find
\begin{align}
 \left(\frac{z_{12}z_{2\bar{1}}}{z_{12}z_{\bar{1}\bar{2}}z_{1\bar{1}}z_{2\bar{2}}}\right)^{2h} \approx \Bigg\{ \begin{split}
     (\gamma)^{4h} e^{-4\gamma h\theta} \; \text{for} \; s>\frac{\theta}{2} \\
     (\gamma)^{4h}e^{-8\gamma hs} \; \text{for} \; s<\frac{\theta}{2}
 \end{split}  
\end{align}
While $F(1)=1$, at $\eta \rightarrow 0$, $F(\eta) \sim (a_{\mathcal{O}})^{2}\eta^{2h_{b}}$. Here, $h_{b}$ is the boundary operator dimension which appears in the leading term of the bulk-boundary OPE of $\mathcal{O}(z,\bar{z})$. Finally putting all these results, we obtain the following expression for two point function at late time limit:
\begin{align}
     {}_{in}\langle \psi|\mathcal{O}(s_{E},0)\mathcal{O}(s_{E},\theta)|\psi\rangle_{in} \sim  \Bigg\{ \begin{split}
         &\gamma^{4h} e^{-4\gamma h\theta} \; \text{for} \; s>\frac{\theta}{2}  \\
         &\gamma^{4h}a_{\mathcal{O}}^{2}e^{8\gamma h_{b}(s-\frac{\theta}{2})}e^{-8\gamma hs} \; \text{for} \; s<\frac{\theta}{2}
     \end{split}
\end{align}
Hence we can see the late time behavior of the two point correlator for $s>\theta/2$ becomes constant, while for $s<\theta/2$ it is decaying exponentially. Similarly, one can compute entanglement entropy by twist operator two point function on a constant $s$ slice and get \footnote{In the following expression $\#$ refers to some constant.}
\begin{align}
    S_{EE}(s) \sim \Bigg\{ \begin{split}
        &\frac{c}{3}\ln\frac{1}{2\gamma} +(\#).s \; \text{for} \; s<\frac{\theta}{2} \\
        & \frac{c}{3}\ln\frac{1}{2\gamma} +(\#).\theta \; \text{for} \; s>\frac{\theta}{2}
    \end{split}
\end{align}
Hence again the late time entanglement entropy shows steady state values as in the global quench scenario\cite{Calabrese:2016xau}. In this way, we can see \textit{the effective dilaton dynamics in the (past) Milne wedge, governed by the proposed Milne vacuum $|\psi\rangle_{in}$ is fundamentally unstable and approaches to a steady state at a late but finite time}. This resembles with the universal feature of global quantum critical quench.

\section{Deformed CFT, static patch and observer's entropy}\label{sec5}
%Match the deformed CFT partition function with that obtained in section 2.
In Lorentzian time, the static patch and Milne wedge provide the same description \cite{Maldacena:2019cbz} with different realization of isometries and coordinates representation related by analytic continuation as in (\ref{Milne to static Lor}). We can try to do similar continuation in Euclidean time. For instance, we analytically continue $s_{E} \rightarrow i\theta'+\frac{\pi}{2L}$ and $\theta\rightarrow it'$ such that
\begin{align}
    ds^{2}=-\frac{ds_{E}^{2}+d\theta^{2}}{\sin^{2}{\frac{s_{E}}{L}}} \rightarrow \frac{d\theta'^{2}+dt'^{2}}{\cosh^{2}{\frac{\theta'}{L}}}
\end{align}
Here $\theta':(-\infty,\infty)$. Upon Lorentzian continuation $t' \rightarrow i\tau$, we will end up with the familiar static patch metric in conformal coordinates as in (\ref{static in conformal})\footnote{We need to identify $L=r_{N}$.}:
\begin{align}
    ds^{2} = \frac{-d\tau^{2}+d\theta^{2}}{\cosh^{2}{\frac{\theta}{L}}}
\end{align}
Here we change the notation $\theta'$ to $\theta$ for this section to avoid excessive notational clutter.  

However doing such analytic continuation $s_{E} \rightarrow s'_{E} + \frac{\pi}{2L}$ followed by $s'_{E} \rightarrow i\theta$, is not possible in the strip picture. The Euclidean time boundary at $s_{E}=0,\frac{\pi}{L}$ breaks the time translational invariance which prevents such analytic continuation. However, we know in static patch has explicit time translational isometry and the Euclidean time should be periodic to remove the conical singularity at the horizons. This problem can be circumvented if we go from strip to ring or cylinder with the same Hamiltonian $H_{c}$ of the form (\ref{defomred Ham})\footnote{One may alternatively think of this like joining another strip to make a cylinder with changing the boundary condition from Dirichlet to periodic. Heuristically, this is just adding the upper Milne wedge and then analytically continue to extended static patch.}. Since the analytic continuation does not change the map from full complex plane to $(\theta,t')$ $z \rightarrow \omega$, the form of the Hamiltonian will not change on the ring. Instead of semicircles, the Euclidean time curve will now become a full circle manifesting time translation symmetry(see fig(\ref{cylinder}). This motivates us to study deformed CFT with Hamiltonian of the form (\ref{defomred Ham}) on a ring. For this section, we will use the same notation for $L_{n},\bar{L}_{n}$ which describes Virasoro modes on the full complex plane.

\subsection{Deformed CFT on a ring, emergent Virasoro algebra and stretched horizon in static patch}\label{sec4.1}
The deformed Hamiltonian (\ref{defomred Ham}) on full complex plane can be written as\footnote{Here we ignore the Casimir term appearing due to the conformal mapping from cylinder to plane.}
\begin{align}
    H_{c} = \frac{i\gamma}{2\pi i}\oint dz \; (z+1)(z-1)T(z) - \frac{i\gamma}{2\pi i}\oint d\bar{z} \; (\bar{z}+1)(\bar{z}-1)\bar{T}(\bar{z})
\end{align}
Using $z=-\coth{\gamma\omega}$ and $\bar{z}=-\coth{\gamma\bar{\omega}}$, with $\omega=\theta+it'$,$\bar{\omega}=\theta-it'$ and the transformation of stress tensor $T(\omega) = (\frac{\partial z}{\partial \omega})^{2}T(z)+\frac{c}{12}Sch\{z,\omega\}$, we end up with:
\begin{align}
    H_{c} &= \frac{1}{2\pi}\left[\int^{\infty}_{-\infty}d\theta\; T(\omega)- \int^{-\infty}_{\infty} d\theta \;\bar{T}(\bar{\omega})-\frac{c}{12}\left(\int^{\infty}_{-\infty}d\theta \;Sch\{z,\omega\}-\int^{-\infty}_{\infty}d\theta \;Sch\{\bar{z},\bar{\omega}\}\right)\right] \nonumber \\
    & = \frac{1}{2\pi}\int^{\infty}_{-\infty} d\theta \; (T(\omega)+\bar{T}(\bar{\omega})) + \frac{c\gamma^{2}}{6\pi} \int^{\infty}_{-\infty}d\theta 
\end{align}
Here we have used constant $t'$ curve as the contour of integration upon mapping from constant time circles in the complex plane. We can see the Schwarzian term will be diverging and we need to cut-off the $\theta$ integral as in \cite{Das:2025cuq}. We choose $\gamma\theta:(-\Lambda_{c},\Lambda_{c})$. The need for cut-off also arises to define Virasoro modes as in \textit{modular quantization} \cite{Tada:2019rls}, \cite{Das:2024mlx},\cite{Das:2024vqe}. After imposing the cut-off the constant term looks similar to a constant \textit{Schwarzian} term appeared as an ADM mass term of JT gravity with negative cosmological constant. As we mentioned in the introduction \ref{sec1}, the ADM clock located at asymptotic boundary plays the role of observer. However, the appearance of this constant Schwarzian term has no clear interpretation to us without a GHY boundary term.

We should also mention that the vacuum of the static patch can be identified with the conformal vacuum on cylinder $|0\rangle_{c}$ such that $H_{c}|0\rangle_{c}=0$. In this vacuum, by using conformal map $z\rightarrow \omega$ and analytic continuation from $t'\rightarrow\tau$, one can obtain two point function ${}_{c}\langle0|\mathcal{O}(\tau,0)\mathcal{O}(0,0)|0\rangle_{c} \propto (\sinh\gamma\tau)^{-2\Delta}$, which shows thermal nature.

One will again observe that this Hamiltonian which should generate the static patch observer time$(t')$, is conserved. We observe
\begin{align}
    \partial_{t'}H_{c} = \frac{1}{2\pi} \left[ T(\infty)-T(-\infty)-\bar{T}(\infty)+\bar{T}(-\infty)\right]
\end{align}
Now $T(\omega) = \left[\frac{1}{\sinh^{4}{\gamma\omega}}T(z)-\frac{c\gamma^{2}}{12\pi}\right]$ and $\bar{T}(\bar{\omega}) = \left[\frac{1}{\sinh^{4}{\gamma\bar{\omega}}}T(z)-\frac{c\gamma^{2}}{12\pi}\right]$. Since $\theta\rightarrow \pm \infty$ corresponds to $z \rightarrow \pm 1$, $T(z)|_{z=\pm 1}$, $\bar{T}(\bar{z})|_{\bar{z}=\pm 1}$ is finite. Thus we can immediately see\footnote{Here again the stress tensor is negative but finite. Even though, one can have thermal two point function and other thermal features of the static patch, this constant negative stress tensor is not concerning. Rather the difference from an positive stress tensor in the AdS$_{2}$ black hole vacuum might be quite encouraging in the sense of distinguished nature of the Nariai black hole. An observer inside the static patch can effectively measure negative local energy density.} %However this static patch vacuum is not a (full) de-Sitter invariant vacuum. Time translation invariance in an expanding universe is just an artifact of forming causal horizons enclosing the observers.}
\begin{align}
    T(\omega)|_{\theta=\pm \infty} = \bar{T}(\bar{\omega})|_{\theta =\pm \infty} = -
    \frac{c\gamma^{2}}{12}
\end{align}
Hence, we have $[T(\omega)-\bar{T}(\bar{\omega})]|_{\theta=\pm \infty} = 0$, that manifests $\partial_{t'}H_{c} = 0$. This is what we dubbed as \textit{emergent conformal boundary condition} at the horizons $\theta=\pm \infty$. We should emphasize that \textit{the fixed points of the deformed Hamiltonian are realized as emergent conformal boundary in the proper time frame of static patch.}

For a free bosonic CFT, the emergent horizon boundary condition can be either a Dirichlet one or Neumann one. Both will give the similar solutionson field modes as we discussed in section \ref{sec2}. After imposing the stretched horizon cut-off we can impose the same condition on the regulator.

%Before constructing the eigenmodes of the Hamiltonian, we compute correlation function as a sanity check. Since the Hamiltonian is constructed out of SL(2,$\mathbb{C}$) generators, the conformal vacuum $|0\rangle_{c}$ of the cylinder left unchanged. Hence we can compute the static patch two point correlator by mapping it to the complex plane and we obtain the following:
%\begin{align}
 %   {}_{c}\langle0|\mathcal{O}(\tau,0)\mathcal{O}(0,0)|0\rangle_{c} \sim (\sinh{\gamma\tau})^{-2\Delta}
%\end{align}
%Here we took scalar operators. This result exactly matches with the static patch retarded propagator

We can now construct \textit{emergent} Virasoro modes as first discussed in the context of dipolar quantization \cite{Ishibashi:2015jba},\cite{Ishibashi:2016bey}\footnote{See also \cite{Das:2020goe}, for a Lorentzian analogue.}. Following \cite{Ishibashi:2016bey}, we define
\begin{align}
    H_{c} = \mathcal{L}_{0}+\bar{\mathcal{L}}_{0}
\end{align}
The corresponding infinitesimal generators are
\begin{align}
    \tilde{l}_{0}=i\gamma(z^{2}-1)\partial_{z}, \; \bar{\tilde{l}}_{0}=-i\gamma(\bar{z}^{2}-1)\partial_{\bar{z}}
\end{align}
Hence the eigenfunction of $\tilde{l}_{0}$ is $f_{k}$ with eigenvalue $k$. 
\begin{align}
    f_{k}(z) = e^{k\int_{z}\frac{dz}{i\gamma(z^{2}-1)}} = \left(\frac{z-1}{z+1}\right)^{-\frac{ik}{2\gamma}}
\end{align}
Thus we can find the infinitesimal modes $\tilde{l}_{k}=i\gamma(z^{2}-1)f_{k}(z)$ and the corresponding conserved charges defined as
\begin{align}
    \mathcal{L}_{k} = \frac{\gamma}{2\pi}\int_{C}dz \; (z-1)^{-\frac{ik}{2\gamma}+1}(z+1)^{\frac{ik}{2\gamma}+1} T(z)
\end{align}
Here $C$ refers to the constant $t'$ contour which denotes a $\theta$ integral over $(\infty,-\infty)$. To construct the commutators of $\mathcal{L}_{k}$, one again need to specify a cut-off at $z=\pm 1+\epsilon e^{i\theta_{\epsilon}}$.

\begin{figure}
    \centering

\begin{tikzpicture}[scale=2.5]

% Axes
\draw[->] (-2,0) -- (2,0) node[right] {$t=\operatorname{Re}(z)$};
\draw[->] (0,-1.6) -- (0,1.6) node[above] {$x=\operatorname{Im}(z)$};

% Labels -1 and +1
\node at (-1,-0.08) {$-1$};
\node at (1,-0.08) {$+1$};
\node at (0.4,1.1) {$\theta$ curve};
\node at (-1.3,0.4) {$t'$ curves};
% Large semicircle from -1 to +1 (center 0)
\draw[thick,->] (-1,0) arc (180:0:1);
%\draw[thick,->] (-1,0) arc (180:0:1);
% Left side semicircles centered at -1
\draw[->] (-0.9,0) arc (-360:0:0.13);
\draw[->] (-0.8,0) arc (-360:0:0.25);
\draw[->] (-0.72,0) arc (-360:0:0.35); % big one reaches to 0

% Right side semicircles centered at +1
\draw[->] (1.1,0) arc (0:360:0.13);
\draw[->] (1.2,0) arc (0:360:0.25);
\draw[->] (1.28,0) arc (0:360:0.35); % big one reaches to 0

\end{tikzpicture}
\caption{Curves of $\theta$ and $t'$ on the full complex plane}
    \label{cylinder}
\end{figure}
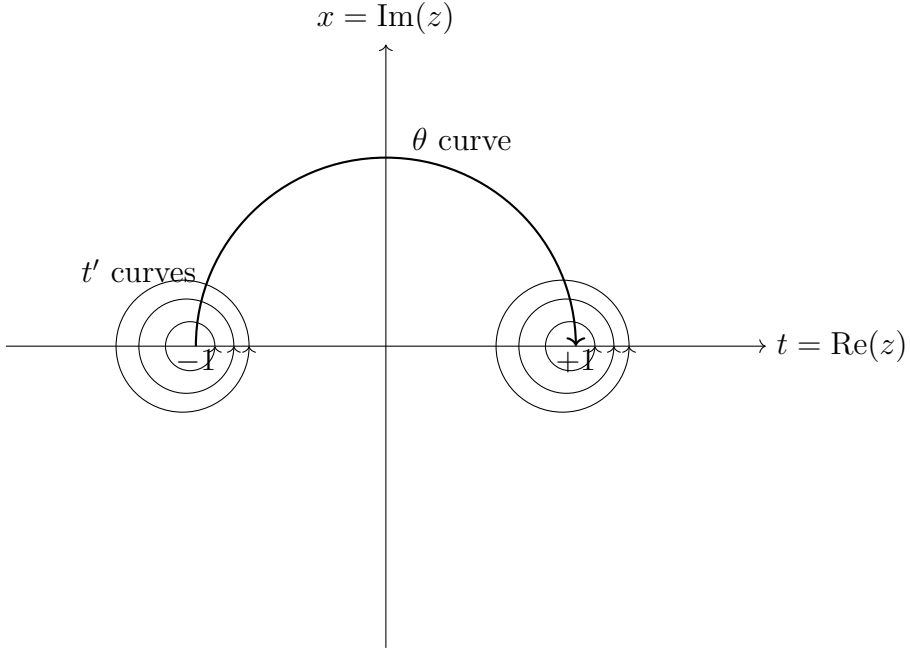

Since the contour of the $(t'-\theta)$
 as in fig(\ref{cylinder}), is just the $\pi/2$ rotated version of the contour of modular quantization in cylinder \cite{Das:2024mlx}, one can check the end result would be similar to that one. We avoid the repetitive computation and we refer to \cite{Das:2024mlx},\cite{Das:2025cuq} for the details. We finally obtain:
 \begin{align}
     [\mathcal{L}_{k}^{\Lambda_{c}},\mathcal{L}_{k'}^{\Lambda_{c}}] = (k-k')\mathcal{L}_{k+k'}^{\Lambda_{c}} + \frac{c_{eff}}{12}(k^{3}+4\gamma^{2}k)\delta_{k+k'}
 \end{align}
$\mathcal{L}_{k}^{\Lambda_{c}}$ is the modified generator having the contour length of $2\Lambda_{c}\equiv\frac{2}{\gamma}\log\left(\frac{2}{\epsilon}\right)$ and $\theta:(-\Lambda_{c},\Lambda_{c})$. Here $c_{eff} = \frac{c}{2\pi\gamma}\ln\left(\frac{2}{\epsilon}\right)$. Here again $k$ is discretized and of the similar form for stretched horizon normal modes obtained in (\ref{normal modes}):
\begin{align}
    k=\frac{2n\pi\gamma}{\ln\left(\frac{2}{\epsilon}\right)}
\end{align}
Here $\gamma=\frac{1}{r_{N}}$ as we identified from the conformal map. Similarly we obtain Virasoro algebra for $\mathcal{\bar{L}}_{k}$ with the similar form. In $\epsilon \rightarrow 0$ limit, one will end up with a continuous Virasoro algebra with dirac delta function as pointed out in \cite{Das:2024mlx}. However, due to conformal boundary condition at $z=\pm 1$, we will end up with a single independent Virasoro algebra. In fact, we can construct the single Virasoro modes with a closed contour with punctures. However the details are not needed for the rest.

\subsection{Modular quantization on static patch and thermal partition function}\label{sec4.2}
To quantize the Hamiltonian we can again use the usual \textit{modular quantization} technique developed in \cite{Das:2024mlx},\cite{Das:2025cuq}. 
Again we are not repeating the procedure. We summarize the key points and some important clarifications from \cite{Das:2025cuq}:
\begin{itemize}
    \item Due to cut-off with conformal boundary condition, the  quantization contour of $(t',\theta)$ curves can be mapped to a thermal BCFT on an annulus of width $W=2\pi \Lambda_{c}$. 
    \item We can construct eigenstates of regulated Hamiltonian $(\mathcal{L}_{0}^{\Lambda_{c}}+\bar{\mathcal{L}}_{0}^{\Lambda_{c}})$ which are localized at the cut-off surface or stretched horizon. These eigenstates are constructed out of primaries as well as modular Virasoro descendants. In \cite{Das:2024mlx},\cite{Das:2025cuq} it is argued that to quantize or to construct the highest weight representation of the modular Virasoro algebra, one need to \textit{uncompactify} the Euclidean time circle, which defines a vacuum and highest weight state using standard path integral definition. However, we now think, this is \textit{not} necessary. \textit{One can define all eigenstates of the cut-off Hamiltonian i.e. primaries and descendants (by choosing constant $s=0$ contour) using $T\mathcal{O}$ ope as described in \cite{Das:2025cuq}. The vacuum can be constructed just by demanding $(\mathcal{L}_{0}^{\Lambda_{c}}+\mathcal{\bar{L}}^{\Lambda_{c}}_{0})|0\rangle_{\Lambda_{c}}=0$.} %In the limit $\epsilon \rightarrow 0$, only conformal primaries may survive located at the fixed points. 
    We note that, these eigenstates are \textit{emergent due to the emergent modular Virasoro algebra and the conformal boundary condition at the cut-off}. All the states at the cut-off surface can be constructed explicitly for massless free bosons  as states of regulated Fock space, briefly described in section \ref{sec2}.
    \item Using this quantization, the $\epsilon \rightarrow 0$ limit can be identified with the zero modes of the Hamiltonian as we will show now. However, \textit{mapping back to $(\omega,\bar{\omega})$ plane these primaries are vanishing, leaving only vacuum at the horizon}\footnote{We took a different claim for these than our earlier paper \cite{Das:2025cuq}, which we think is wrong.}. Hence the vanishing edge modes should further give a vanishing micorcanonical entropy count sourced purely from the surviving vacuum state\footnote{One can not obtain this just naively putting $\epsilon=0$ for which the leading part of entropy is diverging. This different results further suggests that measuring entropy without a cut-off with boundary condition is \textit{meaningless}.}. This on the other hand tells, the entropy explicitly measures the \textit{stretched horizon observer entropy}, which can not be obtained in the semiclassical limit without an observer. The same applies in AdS$_{2}$, where one must impose a cutoff on the dilatons, which can be made arbitrarily large but \textit{cannot be taken literally to infinity.}  
\end{itemize}
Following \cite{Das:2024mlx}, we consider states $|\tilde{0}\rangle$ of the form
\begin{align}
    |\tilde{0}\rangle = \sum_{n}\sum_{m}a_{n}b_{m}L_{-1}^{n}\bar{L}_{-1}^{m}\mathcal{O}_{h}(0,0)|0\rangle_{c}
\end{align}
We want to construct those satisfying
\begin{align}
    H_{c}|\tilde{0}\rangle = 0
\end{align}
Using the identities:
    \begin{align}
    &[L_{1},L^{n}_{-1}] = n(n-1)L^{n-1}_{-1} + 2nL^{n-1}_{-1}L_{0} \nonumber \\
    & [L_{0},L_{-1}^{n}] = n L^{n}_{-1},
\end{align}
we obtain
\begin{align}
    &i\gamma(L_{1}-L_{-1})\sum_{n}a_{n}L_{-1}^{n}\mathcal{O}_{h}(0,0)|0\rangle_{c} \nonumber \\
   & = \sum_{n}i\gamma\left(a_{n+1}n(n+1)+2(n+1)a_{n+1}h-a_{n-1}\right)L^{n}_{-1}\mathcal{O}_{h}(0,0)|0\rangle_{c}
\end{align}
Similarly
\begin{align}
    &i\gamma(\bar{L}_{1}-\bar{L}_{-1})\sum_{m}b_{n}\bar{L}_{-1}^{m}\mathcal{O}_{h}(0,0)|0\rangle_{c} \nonumber \\
   & = \sum_{m}i\gamma\left(a_{m+1}m(m+1)+2(m+1)b_{m+1}h-b_{m-1}\right)\bar{L}^{m}_{-1}\mathcal{O}_{h}(0,0)|0\rangle_{c}
\end{align}
Hence combining this and choosing $a_{n}=\frac{a^{n}}{n!}$, $b_{m}=\frac{b^{m}}{m!}$, we have
\begin{align}
   &\left[ i\gamma(L_{1}-L_{-1})-i\gamma(\bar{L}_{1}-\bar{L}_{-1})\right] |\tilde{0}\rangle \nonumber \\
   &=i\gamma\sum_{n,m}\frac{a^{n}b^{m}}{n! m!}\left[an+2ah-\frac{n}{a}-bm-2bh+\frac{m}{b}\right]L^{n}_{-1}\bar{L}^{m}_{-1}\mathcal{O}_{h}(0,0)|0\rangle_{c}
\end{align}
We can readily see that when $a=b=1$ or $a=b=-1$, we get the desired result i.e. $H_{c}|\tilde{0}\rangle = 0$. Hence the zero-mode vacuum state becomes
\begin{align}
    |\tilde{0}\rangle=\Bigg\{\begin{split} 
     &e^{L_{-1}}e^{\bar{L}_{-1}}\mathcal{O}_{h}(0,0)|0\rangle_{c} = \mathcal{O}_{h}(1,1)|0\rangle_{c} \\
     & e^{-L_{-1}}e^{-\bar{L}_{-1}}\mathcal{O}_{h}(0,0)|0\rangle_{c} = \mathcal{O}_{h}(-1,-1)|0\rangle_{c}
     \end{split}
\end{align}
These are precisely the primaries $\mathcal{O}(z,\bar{z})$ located at the fixed points $z,\bar{z}=\pm 1$. However, mapping into the $\omega,\bar{\omega}$ plane, we can see $\mathcal{O}(\omega,\bar{\omega})|_{\theta=\pm \infty}$ is vanishing due the Jacobian factor of the conformal mapping for primaries.
\begin{align}
   \left[\left(\frac{dz}{d\omega}\right)^{h}\left(\frac{d\bar{z}}{d\bar{\omega}}\right)^{h}\right]_{\theta=\pm\infty}=0 
\end{align}
Hence only \textit{vacuum or identity operator} survives exactly at the $\epsilon\rightarrow 0$ limit. %However, the zero mode primaries in the $\epsilon \rightarrow 0$ limit is precisely appeared due to \textit{the emergent boundary condition at the stretched horizon cut-off}. These are the sole effect of the modular quantization and modular Virasoro algebra. 
This again shows \textit{the Hilbert space of modular quantization is completely different and emergent due to the boundary condition at near horizon cut-off}. %Hence the thermal entropy sourced from this quantization is an artifact of \textit{counting high energy eigenstates of the cut-off Hamiltonian as a residual effect of emergent stretched horizon boundary condition}.

%We will end this section by a brief outline of computing thermal entropy in modular quantization.
To compute the thermal entropy of the annulus we again follow the closed-open string duality \cite{Cardy:2016fqc}. The annulus partition function of temperature $\frac{4\pi^{2}}{W}$ can be obtained using closed string channel partition function in the annulus of length $W=2\nu\Lambda_{c}=2\log(\frac{2}{\epsilon})$ as
\begin{align}
    Z_{\epsilon}={}_{\epsilon}\langle B|e^{-W(L_{0}+\bar{L}_{0}-\frac{c}{12})}|B\rangle_{\epsilon}
\end{align}
In $\epsilon \rightarrow 0$, similarly to \cite{Das:2024mlx}, we end up with
\begin{align}
    S_{\epsilon \rightarrow 0}= \frac{c}{3}\log(\frac{2}{\epsilon})+g_{a}+g_{b}
\end{align}
Here $g_{a,b}\equiv \log{\langle0|h\rangle\rangle}_{a,b}$ are the boundary entropies. Since in $\epsilon \rightarrow 0$ limit, $|h\rangle\rangle \rightarrow |0\rangle$, $g_{a,b}$ vanishes identically. Hence the leading term matches with the entanglement entropy in a full time slice of the static patch $\theta:(-\Lambda_{c},\Lambda_{c})$ as one can check by computing twist operator two point function using the map $z\rightarrow \omega$. This is again consistent with the fact that the time translation of static patch Hamiltonian is identified with the flow by \textit{modular Hamiltonian} on the full patch bounded by the horizons.

On the other hand, we can compare this result by explicit computation of free massless bosons($c=1$) in the stretched horizon background as in (\ref{out entropy}). We have seen a \textit{potential mismatch} of overall factor of $\frac{1}{8}$. %Naively, If we have 8 massless scalars in the static patch background, this would exactly match with a single scalar in the worldsheet description. 
Currently we have no clear understanding behind this mismatch. Neglecting that potential mismatch, we can see both of these terms proportional to $\log(\frac{2}{\epsilon})$\footnote{Here we took $r_{N}=1/\nu=1$. However for the general $r_{N}\neq 0$, we may need to consider $H_{c} = \alpha L_{0}+i\gamma(L_{1}-L_{-1})$ with $d=\alpha^{2}-4\gamma^{2}<0$. This class of Hamiltonian belongs to the \textit{`heating phase'} in the Floquet CFT language \cite{Han:2020kwp},\cite{Das:2024vqe}.}. 

We end this section by commenting our take on the questions we have tried to address \textit{iff we may think of this deformed CFT quantization as an UV description of near Nariai black hole or extended static patch}:
\begin{itemize}
    \item[1.] \textit{What sources the entropy of static patch in the near Nariai reduction to two dimension?}

    Thermal entropy of free massless bosons in stretched horizon background. Along the line of \cite{Das:2024vqe}, one can again think the same as the entanglement entropy of the causal diamond vacuum with the \textit{outside}\footnote{This is exactly the other side of the quantization contour as described in \cite{Das:2024vqe}.}. The stretched horizons approaching the Naiai horizons or $\epsilon\rightarrow 0$ limits are the \textit{semi-classical limits} in this description. %However, we \textit{can not} take exactly $\epsilon = 0$.

\item[2.] \textit{What is the (Nariai) static patch holography in this context?}

All the (dilaton) entropic degrees of freedom of the extended static patch is restored in the stretched horizon and the Hilbert space is defined by the modular quantization of worldsheet CFT.

    \item[2.] \textit{Who is the observer in the story?}

    The stretched horizon or the moving membrane defined by the worldsheet CFT with deformed Hamiltonian on a ring, plays the role of a gravitating observer, as observed by a (non-gravitating) podal observer located at the center of the static patch.
\end{itemize}
\section{Discussion}\label{sec6}
In this note, our main goal is to find possible UV description of effective 2D theory of dilatons arising from reduction of higher dimensional Einstein gravity describing pure and black holes in dS. While a stable dS$_{2}$ vacuum is difficult to find in any time-dependent patches of dS$_{2}$, we have managed to describe a sensible description of (Nariai) static patch effective theory in terms of worldsheet CFT with deformed Hamiltonian. Modular quantization of the worldsheet CFT further introduces a stretched horizon with an emergent conformal boundary condition, which acts like an internal observer in this description. In the CFT description, we have not taken $L \rightarrow iL$, since the role reversal of space-time direction in the proper time frame of the deformed Hamiltonian already manifests a CFT on 2d dS metric upto an \textit{overall change in Lorentzian signature}. To be more precise, the effective metric defined by the deformed CFT on strip has the following form upon analytic continuation in Lorentzian time:
\begin{align}
    ds^{2}_{deformed}= 4L^{2}_{AdS}\gamma^{2}\frac{ds^{2}-d\theta^{2}}{\sinh^{2}{2\gamma\theta}} = - \left(4L^{2}_{AdS}\gamma^{2}\frac{-ds^{2}+d\theta^{2}}{\sinh^{2}{2\gamma\theta}}\right)
\end{align}
If we do not ignore the overall negative sign of the metric, one may interpret it as AdS$_{2}$ Rindler patch with $\theta$ plays the role of time. On the other hand, if we could ignore the overall sign, we should think the same as Milne patch in de-Sitter\footnote{Note that, if we wick-rotate the UHP coordinates $z\rightarrow iz$ and $\bar{z}\rightarrow i\bar{z}$, we end up with right half plane(RHP). The metric in the RHP is EAdS$_{2}$ Poincare with an overall negative sign, which alternatively could be obtained by $L \rightarrow iL$ without wick-rotating the coordinates. Hence we may think of going UHP to RHP is same as changing $L \rightarrow iL$.}. In the latter case, one may alternatively think of this set-up as de-Sitter \textit{mimicker}, rather than an inherently de-Sitter due to negative cosmological constant\footnote{This also justifies a negative stress tensor.}. However, we would like to think of these deformed Hamiltonians of CFT generates 2d CFT on de-Sitter patches, which, to the best of our knowledge, is not discussed previously in the literature. It would be definitely interesting to find a set-up where negatively divergent energy density in our proposed Milne vacuum could be compensated by gluing it with a very high positive energy density state like large AdS$_{2}$ blackhole. This could probably lead to some version of cosmology in open universe. 

We now discuss some further important points and possible future direction we gained from this note.
\begin{itemize}
  %  \item \textit{Exact UV completion and string theory embedding:} The next and ultimate goal of this program is to embed this deformed worldsheet CFT Hamiltonians in a concrete UV complete quantum gravity theory like string theory. Without this, the program will still be incomplete and makes no sensible direction and motivation. Even though, we have a lack of present understanding on this, We will hope to return with some direction in future. 

\item\textit{Near horizon universality and deformed CFTs:} At this point of our survey, we can conjecture about the emergence of a universal conformal symmetry in the near-horizon limit for a broad class of black holes. This symmetry can be realized as a free bosonic CFT in 2D Rindler space for many non-extremal flat and AdS black holes \cite{Solodukhin:1998tc} \footnote{A more careful treatment of boundary contributions, including potential GHY terms, is still required.}; as a free bosonic CFT in AdS$_2$ supplemented by a boundary Schwarzian term for near-extremal flat and AdS black holes; and as a free bosonic CFT in dS$_2$ for Nariai black holes near the horizon. \textit{Unlike the situation in JT gravity}, this universal free bosonic CFT offers a more direct connection to worldsheet CFTs with SL(2,$\mathbb{R}$) deformed Hamiltonians\footnote{We must clarify that the name \textit{deformed CFT} should not be taken as deforming conformal symmetry.}. A more thorough investigation and literature survey of the near-horizon limits of larger class of higher dimensional as well as supersymmetric black holes may yield a deeper understanding of this near horizon quantum gravity/ deformed CFT correspondence. We think that the most important task is to embed this sort of description in a concrete UV complete framework like superstring theory.

\item\textit{Source of extremal entropy from deformed CFTs:}
Even though, the worldsheet approach for effective dilaton theory could give us an understanding of near extremal (and near horizon) entropy in terms of entanglement entropy, the source for pure extremal entropy or Nariai entropy is not clear to us from the same worldsheet description. An external charged sector of the CFT may provide a hint towards the counting \cite{Gupta:2008ki}, \cite{Almheiri:2016fws}, \cite{Poojary:2022meo}.  %The near horizon limit of extremal black holes have universal SL(2,$\mathbb{R}$)$\times U(1)$ isometry. From our worldsheet analysis, we can say the SL(2,$\mathbb{R}$) deformed Hilbert space would provide the source of the near horizon near extremal entropy. The other $U(1)$ symmetry was speculated to be a part of external $U(1)$ conserved charge\cite{Das:2025cuq} of the worldsheet theory. This is partly justified due to the identification of dual holographic deformed CFT Hilbert space to pure JT gravity Hilbert space(without the topological part) as in \cite{Das:2025cuq}. The identification was based on the dimensional reduction of planar black holes(with a cut-off) in pure 3d Einstein gravity to JT gravity. If one use the similar reduction for the near extremal planar BTZ , one would end up with an additional extremal sector. Hence the $U(1)$ symmetry should be associated to dual holographic CFT with an additional or external charge sector \cite{Gupta:2008ki}. We should emphasize that, the holographic CFT dual to pure Einstein gravity is made out of only stress tensor sector and hence does not contain any primaries. However, for the dimensional reduction of $d\geq 4$ near extremal black holes as described in the appendix (\ref{A}), we now think this $U(1)$ symmetry should correspond to an internal one of the worldsheet theory. We note that, the deformed Hamiltonians have a huge(in principle , infinite) degenerate spectrum of zero modes as observed in \cite{Das:2024mlx},\cite{Das:2025cuq} and also in the present paper. For a massless free bosons, these are the vertex primaries located on the fixed points of the form $:e^{ib\phi(z_{f})}:$ with dimensions $\Delta \propto b^{2}$. Hence these zero mode operators can serve as an $U(1)$ symmetry of the deformed Hamiltonians. Notably, as we clarified in this paper, these zero modes would not appear in the modular quantized Hilbert space. Correspondingly, the full Hilbert space can be thought of as $\mathcal{H}_{full} = \mathcal{H}_{zero}\otimes \mathcal{H}_{mod}$, where $\mathcal{H}_{zero}$ is constructed out of the zero modes spectrum located at the fixed point of the worldsheet Hamiltonian; whereas $\mathcal{H}_{mod}$ is constructed out of modular quantization with modular Virasoro algebra. We speculate the proper counting of degenerate zero modes \footnote{The full UV description should be needed to get a finite entropy counting from such infinite degeneracy.} could give rise to the entropy of extremal sector arising from dimensional reduction of higher dimensional$(d\geq 4)$ near extremal black holes.
\end{itemize}
We end our discussion with some remarks on the emergence of different worldsheet times. %Serious readers may skip this part.
\subsection{Comment on emergence of observer time} 
In this note, we only focus our attention to past Milne wedge and extended static patch of a full dS$_{2}$. The failure of constructing a stable Milne vacuum is circumvented by analyzing a sensible static patch worldsheet description with a deformed CFT on a ring. One may wonder, whether an effective dilaton theory in global dS$_{2}$ or in a flat slicing could be similarly described. %We expect the corresponding worldsheet deformed Hamiltonian might be of the form $\alpha L_{0}+i\gamma(L_{1}-L_{-1})$. One can check, the proper time curves for the Hamiltonians with $\alpha^{2}>4\gamma^{2}$ and $\alpha^{2}=4\gamma^{2}$ describes global and flat slicing of dS$_{2}$ respectively. 
At present, it is not clear to us on how to construct the vacuum for them respectively\footnote{There could be issues involving the antipodal identification, which may suggest the deformed worldsheet theories for those geometries may not be directly correspond to a strip or cylinder.}. In \cite{Das:2025cuq}, we have  noticed that Hamiltonians $\alpha L_{0}+\beta(L_{1}+L_{-1})$ with $\alpha^{2}>4\beta^{2}$, $\alpha^{2}=4\beta^{2}$ and $\alpha^{2}<4\beta^{2}$ would describe AdS$_{2}$ in global, Poincare and black hole coordinates. Hence, more generally one can consider SL(2,$\mathbb{R}$) deformed Hamiltonians on a ring of the form
\begin{align}
    H_{F}=\alpha L_{0}+\beta(L_{1}+L_{-1})+i\gamma(l_{1}-L_{-1})+cc
\end{align}
Note that, taking $\gamma=0$ and $\alpha^{2}=4\beta^{2}$  would describe CFT in AdS$_{2}$ Poincare(cold black hole), $\beta=0$ and $\alpha^{2}<4\gamma^{2}$ would describe CFT in $dS_{2}$ static patch and $\alpha=\gamma=0$ will describe modular Hamiltonian of a CFT on an interval of a ring\cite{Das:2024vqe}. It is well known that under (global) conformal transformation the modular flow generated by Rindler Hamiltonian is mapped to a spherical domain of a Lorentzian CFT\cite{Casini:2011kv}. %However, if $\alpha$ is \textit{imaginary}, then for $\beta=\gamma=0$, the CFT will be described on \textit{Rindler spacetime}. However that makes the Hamiltonian to be anti Hermitian. 
For general $\alpha,\beta,\gamma \neq 0$, it would describe some complex metric which is not understood at this stage. %signifying out-of-equilibrium or unstable situation
Now we consider a slightly modified class of deformed Hamiltonian of a CFT on a ring of the form:
\begin{align}\label{Floquet Ham}
    H_{F}=i\alpha L_{0}+\beta(L_{1}+L_{-1})+i\gamma(l_{1}-L_{-1})+cc
\end{align}
The negative sign of the SL(2,$\mathbb{R}$) Casimir suggests it describes a \textit{heating phase} in the Floquet CFT language \cite{Wen:2018agb}-\cite{Das:2022jrr}. Taking $\alpha=\beta=0$ describes de-Sitter static patch, whereas $\alpha=\gamma=0$ describes AdS$_{2}$ Rindler. On the other hand, $\beta=\gamma=0$ will describe CFT$_{2}$ on 2d Rindler or Minkowski spacetime\cite{Mao:2025cfl} \footnote{Depending on the positivity or negativity of the energy density on positive or negative real axis, one may treat that as either Rindler or Minkowski Hamiltonian.}. The non-Hermiticity of the Hamiltonian may be interpreted as post-selected weak measurement as in \cite{Lapierre:2025zsg}. However, we will show momentarily how this apparent non-unitarity emerges from an \textit{inherently Unitary dynamical set-up}.

Notably, \textit{a charged SdS$_{4}$ black-hole is known to possess three degenerate extremal limits: dS$_{2}\times$S$_{2}$, AdS$_{2}\times$S$_{2}$ and Mink$_{2}\times$S$_{2}$\cite{Romans:1991nq},\cite{Mann:1995vb} \footnote{In fact, for KN SdS$_{2}$ black holes, there exist two `ultracold' limits consisting Rind$_{2}\times$S$_{2}$ and Mink$_{2}\times$S$_{2}$\cite{Booth:1998gf}. However, CFT in Mink$_{2}$ may be related to Euclidean angular quantization of Rindler Hamiltonian\cite{Agia:2022srj},\cite{Agia:2024wxx}. A similar discussion has also been appeared in \cite{Das:2024mlx}, where it was dubbed as `Lorentzian representation of Euclidean quantization'.}. These arise respectively from taking the Nariai limit, the limit in which the outer and inner horizon radii coincide, and the limit in which all three horizons (outer, inner, and cosmological) become degenerate.}\footnote{The author thanks Debangshu Mukherjee \cite{Maulik:2025phe} for pointing out the reference \cite{Castro:2022cuo}.}. A natural question then which of these degenerate limits correspond to a stable equilibrium situation\cite{Shi:2025amq}-\cite{Aalsma:2025lcb}. To find the effective classical theories in those limits, one may follow the S-wave reduction of Einstein Maxwell theory in 4D with positive cosmological constant similar to the spirit as described in appendix \ref{A} with or without adding a potential GHY term. In any case, we should expect to obtain free massless dilatons in those limiting 2D geometries with a worldsheet volume terms and boundary terms.

However, the point we wish to emphasize is that, from a UV perspective, the different choices of times (corresponding to different Hamiltonians) should not be \textit{ a priori} preselected, but rather be \textit{emergent}. In the periodically driven CFT setup\cite{Wen:2018agb}-\cite{Das:2022jrr}, such Hamiltonians appear \textit{naturally}. For instance, in the simplest scenario one can consider a two-step drive protocol, in which a CFT on a ring is evolved with a Hamiltonian $H_{1}$ for a time interval $T_{1}$, followed by another Hamiltonian $H_{2}$ for a time interval $T_{2}$. This defines an inherently time-dependent system under drive. One can then apply this periodic driving $n$ times, as in:

\begin{align}
    \underbrace{ e^{-iH_{1}T_{1}}e^{-iH_{2}T_{2}}e^{-iH_{1}T_{1}}e^{-iH_{2}T_{2}}\dots e^{-iH_{1}T_{1}}e^{-iH_{2}T_{2}}}_{\text{n times}} = e^{-iH_{F}t}
\end{align}
Here $H_{F}$ is the \textit{time independent} Floquet or effective Hamiltonian and generates a \textit{new stroboscopic time} $t=n(T_{1}+T_{2})$. In this driven CFT set up, one can change different phases \textit{dynamically} by changing the frequency and amplitude of the drive parameter. Hence the question in our context is whether we could \textit{naturally} obtain a Floquet Hamiltonian of the form (\ref{Floquet Ham}) by choosing certain $H_{1}, H_{2}$.

Let us consider $H_{1}=a(L_{1}+L_{-1})$ and $H_{2}=ib(L_{1}-L_{-1})$ with real $(a,b)$. These are Hermitian operators as in radial quantization. By using Baker-Campbell-Hausdorff(BCH) formula we can in principle compute $e^{-iHT_{1}}e^{-iHT_{2}}$ as a nested commutators of $H_{1},H_{2}$ in Euclidean time. One could easily check that, for a single period this would take the form of (\ref{Floquet Ham}) or $i\alpha L_{0}+\beta(L_{1}+L_{-1})+i\gamma(L_{1}-L_{-1})+cc$. Here $\alpha,\beta,\gamma$ depends of $a,b,c,T_{1},T_{2}$. Hence this apparent non-unitarity of the Floquet Hamiltonian arises from a Unitary periodically driven CFT set-up. However, in the Floquet time $t$, one can treat this as a Hermitian Hamiltonian with a different notion of Hermiticity\footnote{The notion of Hermiticity explicitly depends on the choice of quantization time as in \cite{Ishibashi:2016bey},\cite{Das:2024mlx}. The notion of adjoint operation in radial quantization may not coincide with  a different quantization if the Hamiltonians are not unitarily related to each other.}.
%HBy choosing appropriately $H_{1,2}$ one could end up with $H_{F}$ as in (\ref{Floquet Ham}) \cite{Das:2022pez}, \cite{Das:2023xaw},\cite{Banerjee:2024zqb}.  Hence different stable configuration of $\alpha,\beta,\gamma$ should correspond to taking different  phases. A detailed study of stability of those dynamical phases could be useful to understand those degenerate limits of charged SdS or other near extremal limits in AdS or flat black holes. For instance, from the phase diagram of \cite{Banerjee:2024zqb}, one can see the \textit{heating phases} \footnote{This includes $\alpha^{2}<4\gamma^{2}$ as well as $\alpha=\gamma=0$} are separated from distinct non-heating phases by a \textit{phase boundary}\footnote{This includes $\alpha^{2}=4\beta^{2}$,$\gamma=0$.}.

Nevertheless, the main point we want to make here is that \textit{the time $t$ inherent to the Floquet Hamiltonian $H_{F}$, is a stroboscopic time.} In this Floquet sense, the \textit{stroboscopic time is emergent}. To be more precise, the Floquet observer will measure stroboscopic time in their own clock which is completely different than the clocks for the observers associated to $H_{1,2}$. This is analogous to find a static patch in an accelerating or time dependent (dS) universe. Consequently, to account for the cost of residing in an accelerating universe, static observers are required to observe features like surrounding cosmological (and black hole) horizons. We know that Floquet Hamiltonians in heating phases also exhibit features such as \textit{fixed points}. We feel, the gravitational observers measure the stroboscopic Floquet time in the worldsheet description, which is inherently emergent. The principle behind the fine tuning of choosing the Hamiltonians $H_{1,2}$, drive parameters and frequencies still remains elusive to us from a gravitational perspective. %We hope to return with more evidences on this in near future.

\textbf{Acknowledgement:} The author(SD) of this paper strongly believes that any scientific progress is a collective and collaborative
process where the authorship is an effort to consolidate all the ideas in a single manuscript. He would like to thank Bobby Ezhuthachan, Rahul Roy and Krishnendu Sengupta for discussion at the early stage of this project. He also wants to thank collectively Vaibhab Burman, Dibya Chakraborty, Anirban Dinda, Arnab Kundu, Parthasarathi Majumdar, Debangshu Mukherjee and Koushik Ray, as well as ChatGPT and Perplexity for related discussions and help. He is also grateful to the co-authors of \cite{Das:2024vqe}: Bobby Ezhuthachan, Somnath Porey and Baishali Roy for noticing and emphasizing the paper by Solodukhin \cite{Solodukhin:1998tc}, which provides an important clue to understand the reduction process of this note. He would also like to acknowledge the organisers of the school ``Introductory School on Conformal Field Theories" at IACS(18-22 Aug,2025), which helps him to clarify some basic 2D CFT stuff. He would also like to thank the organizers as well as speakers, participants and poster presenters of Indian String Meeting (2025), who help him to clarify many doubts and conceptual mistakes regarding this work. The research work of SD is supported by a DST INSPIRE Faculty Fellowship.

\appendix
\section{JT gravity and deformed CFT from dimensional reduction of near extremal black holes in near horizon limit}\label{A}

In \cite{Das:2025cuq}, we claim that the classical limit of JT gravity coupled to CFT with $\mathcal{O}(1)$ central charge is described by a certain deformed CFT in a prescribed limit. Partial justification of this claim was made by the fact that in the semiclassical limit, the backreaction of the conformal fields can be neglected and the dynamics of JT coupled to matter CFT is boiled down to the dynamics of CFT in the UV frame. We recovered the entropy of the coupled system away from extremality, purely from the entropy of deformed CFT in the prescribed classical limit. Here in this appendix, we provide a concrete set-up for this justification where JT coupled to free bosons with $c=1$ arises naturally from S-wave reduction in the near extremal near horizon limit of 4D RN black holes in AdS.

Unlike in de-Sitter, to obtain AdS$_{2}$ by dimensional reduction, one must need to add angular momentum or electric or magnetic charges in the higher dimensional AdS or flat space black holes. In the extremal and near horizon limit the metric becomes factorized to AdS$_{2}\times$S$^{2}$ \cite{Maldacena:1998uz}. On the other hand, taking near horizon first and then taking near extremal limit would give same factorization with a AdS$_{2}$ black hole geometry\cite{Sen:2008yk}. Near extremal physics is well-known to be described universally by JT gravity. Here we perform an Weyl-equivalent reduction similar to section \ref{sec 2.1} in the same limit. 

Consider the following magnetically charged AdS$_{4}$ black holes with the following metric:
\begin{align}\label{RN bh}
    ds^{2} = -(1-\frac{2G_{4}M}{r}+\frac{4\pi Q_{m}^{2}}{r^{2}}+\frac{r^{2}}{L^{2}})dt^{2}+\frac{dr^{2}}{(1-\frac{2G_{4}M}{r}+\frac{4\pi Q_{m}^{2}}{r^{2}}+\frac{r^{2}}{L^{2}})} + r^{2}d\Omega_{2}^{2}
\end{align}
Here $M,Q_{m}$ are the mass and magnetic charge of the black hole. For the large AdS black hole with $r_{h}>>L$, the extremal charge and mass become \cite{Nayak:2018qej}
\begin{align}
    Q_{ext}^{2} = \frac{3r_{h}^{4}}{4\pi L^{2}} , \; M_{ext} = \frac{2r_{h}^{3}}{G_{4}L^{2}} 
\end{align}
Taking the near horizon limit $r \rightarrow r_{h}$ yields the following \cite{Nayak:2018qej}
\begin{align}\label{factorized RN}
 ds^{2} =  -\frac{(r-r_{h})^{2}}{L_{2}^{2}}dt^{2}+\frac{L_{2}^{2}}{(r-r_{h})^{2}}dr^{2}+r_{h}^{2}d\Omega_{2}^{2} +\mathcal{O}\left(\frac{r-r_{h}}{r_{h}}\right) 
\end{align}
Here, the AdS$_{2}$ radius is $L_{2} = \frac{L}{\sqrt{6}}$. This is also true for flat space black holes in the same limit. To obtain the effective gravitational theory in this limit, we begin with 4D Einstein-Maxwell action with negative cosmological constant $\Lambda=-\frac{3}{L^{2}}$, accompanied with a boundary GHY term:
\begin{align}
    I = \frac{1}{16\pi G_{4}}\int_{M_{4}}d^4 X \sqrt{-\hat{g}_{4}}(\hat{R}_{4}-2\Lambda_{4}-4\pi F^{2}) +\frac{1}{8\pi G_{4}}\int_{\partial M_{4}}d^{3}X \sqrt{-\hat{\gamma}_{3}} K_{3}
\end{align}
Unlike the de-Sitter case, here one needs to add the GHY term where $\hat{\gamma}_{3}$ is the induced metric on the three dimensional asymptotic boundary and $K_{3}$ is the trace of extrinsic curvature of the boundary. We also consider the following class of asymtptotically AdS$_{4}$ metrics with the following form
\begin{align}
    ds_{4}^{2} = g_{\mu\nu}dx^{\mu}dx^{\nu} + r_{h}^{2}\phi(x)d\Omega_{2}(\theta,\chi)^{2}
\end{align}
Further assuming the metric as a solution of Einstein-Maxwell theory, satisfying the equation of motion $F_{\theta\chi}=Q_{m}\sin\theta$, the 4D action reduces to the following 2D action:
\begin{align}
    I=\frac{1}{16\pi G_{2}}\int_{M_{2}}d^{2}x \sqrt{-g_{2}}\left(\phi R_{2}-2\Lambda_{4}\phi+\frac{2}{r_{h}^{2}}+\frac{(\nabla\phi)^{2}}{2\phi}-\frac{8\pi Q_{m}^{2}}{r_{h}^{4}\phi}\right) +\frac{1}{8\pi G_{2}}\int_{\partial M_{2}} \sqrt{-\gamma} K
\end{align}
Here $\frac{1}{G_{2}}\equiv \frac{r_{h}^{2}\Omega_{2}}{G_{4}}$. We note that the total derivative terms from the bulk action gets cancelled from a similar term appearing in the boundary action upon reduction \cite{Svesko:2022txo}. Again by performing a Weyl rescaling $g_{\mu\nu}\rightarrow \omega^{2}(x)\bar{g}_{\mu\nu}$, we end up with
\begin{align}
    I &= \frac{1}{16\pi G_{2}}\int_{\bar{M}_{2}}d^{2}x \sqrt{-\bar{g}_{2}}\left( \phi\bar{R}_{2}+\frac{2}{\omega}(\nabla_{\mu}\phi)(\nabla^{\mu}\omega)-2\Lambda\phi\omega^{2}+\frac{2\omega^{2}}{r_{h}^{2}}+\frac{(\nabla\phi)^{2}}{2\phi}-\frac{8\pi Q_{m}^{2}\omega^{2}}{r_{h}^{4}\phi} \right) \nonumber \\
    & \; \; \; \;+ \frac{1}{8\pi G_{2}}\int_{\partial \bar{M}_{2}} \sqrt{-\bar{\gamma}}\phi_{b} \bar{K}
\end{align}
Once again a total derivative term from the bulk action $2\nabla^{\mu}\left(\frac{\phi}{\omega}\nabla_{\mu}\omega\right)$ is canceled by a similar term in the boundary action under rescaling. By choosing $\omega \propto \frac{1}{\phi}$ and taking the near horizon limit by expanding $\phi=\phi_{0}+\Phi$ one end up with a standard JT action with boundary and topological Einstein-Hilbert term \cite{Nayak:2018qej}. However, as in section \ref{sec 2.1}, we choose $\omega=c\phi^{-\frac{1}{4}}e^{\beta\phi}$ and obtain the following:
\begin{align}
   I &= \frac{1}{16\pi G_{2}}\int_{\bar{M}_{2}}d^{2}x \sqrt{-\bar{g}_{2}}\left( \phi\bar{R}_{2}+2\beta(\nabla\phi)^{2}+2\Lambda c^{2}e^{2\beta\phi}\left(\frac{1}{\sqrt{\phi}}-\sqrt{\phi}\right)-\frac{8\pi Q_{m}^{2}}{r_{h}^{4}}c^{2}\phi^{-\frac{3}{2}}e^{2\beta\phi} \right) \nonumber \\
    & \; \; \; \;+ \frac{1}{8\pi G_{2}}\int_{\partial \bar{M}_{2}} \sqrt{-\bar{\gamma}}\phi_{b} \bar{K}
\end{align}
Expanding $\phi=\phi_{0}+\Phi$ around $\phi_{0}=1$, we have
\begin{align}
  2\Lambda c^{2}e^{2\beta\phi}\left(\frac{1}{\sqrt{\phi}}-\sqrt{\phi}\right) = -2\Lambda c^{2}e^{2\beta}\Phi +\mathcal{O}(\Phi^{2}), \; \text{and}\; c^{2}\phi^{-\frac{3}{2}}e^{2\beta\phi} = c^{2}e^{2\beta}\left[1+(2\beta-\frac{3}{2})\Phi+\mathcal{O}(\Phi^{2})\right]. 
\end{align}
Choosing $\beta=\frac{1}{4}$, $c = e^{-\frac{1}{4}}$ and expanding upto linear order in $\Phi$, the action reduces to
\begin{align}
    I = &\frac{1}{16\pi G_{2}}\int_{\bar{M}_{2}}d^{2}x \sqrt{-\bar{g}_{2}} \left[\frac{1}{2}(\nabla\Phi)^{2}+\Phi\left(\bar{R}_{2}-2\Lambda+\frac{8\pi Q_{m}^{2}}{r_{h}^{4}}\right)\right] + \frac{1}{8\pi G_{2}}\int_{\partial \bar{M}_{2}} \sqrt{-\bar{\gamma}}\Phi_{b} \bar{K}\nonumber \\
    &+\frac{\phi_{0}}{16\pi G_{2}}\left(\int_{\bar{M}_{2}} d^{2}x \sqrt{-\bar{g}_{2}} \bar{R}_{2}+2\int_{\partial \bar{M}_{2}}\sqrt{-\bar{\gamma}}\bar{K}\right) - \frac{\phi_{0}Q_{m}^{2}}{2r_{h}^{4}G_{2}} \int_{\bar{M}_{2}} d^{2}x \sqrt{-\bar{g}_{2}} \\
    \equiv &I_{0}+I_{1}+I_{2}+I_{3}
\end{align}
Now putting the extremal limit or substituting $Q_{m}$ by $Q_{ext}$, we can see 
\begin{align}
    -2\Lambda+\frac{8\pi Q_{ext}^{2}}{r_{h}^{4}} = \frac{12}{L^{2}} = \frac{2}{L_{2}^{2}}=-2\Lambda_{2}
\end{align}
As we argued previously, we have freedom to choose $\bar{g}_{\mu\nu}$ to be locally AdS$_{2}$ which is Weyl transformed to $g_{\mu\nu}$ with our chosen Weyl factor $\omega$. Since we are interested to study the near extremal and near horizon limit of (\ref{RN bh}), we choose $\bar{g}_{\mu\nu}$ to be in the form of (\ref{factorized RN}).
\begin{align}
    ds^{2}_{\bar{g}} = -\frac{(r-r_{h})^{2}}{L_{2}^{2}}dt^{2}+\frac{L_{2}^{2}}{(r-r_{h})^{2}}dr^{2} = -4L_{2}^{2}\frac{ dz_{+} dz_{-}}{(z_{+}-z_{-})^{2}}, \; \text{where} \; z_{\pm}=t\pm L_{2}u, u=\frac{L_{2}}{r-r_{h}}.
\end{align}
For this metric $\bar{R}_{2} = -\frac{2}{L_{2}^{2}}$. Hence $I_{0}$ reduces to free massless scalar field action in the AdS$_{2}$ Poincare background. $I_{2}$ is the topological term $I_{top}$ and it provides the extremal entropy of the 4D black hole. $I_{3}$ is the world-sheet volume term which may be dropped by using suitable counter term\footnote{Note that, this is not required to drop this term. However, here our motivation is just to connect with JT gravity for which this may not be important to us.}. The boundary term $I_{1}$ should also be regulated by using a similar counter term. Hence the final form reduces to
\begin{align}
    I = \frac{1}{16\pi G_{2}}\int_{AdS_{2}}   d^{2}z\sqrt{-g_{\{z_{+},z_{-}\}}} \frac{(\nabla_{\{z_{+},z_{-}\}}\Phi)^{2}}{2}  + \frac{1}{8\pi G_{2}}\int_{\partial AdS_{2}} \sqrt{-\bar{\gamma}}\Phi_{b} (\bar{K}-1) + I_{top}
\end{align}
In this rigid AdS$_{2}$ frame, the boundary term captures the asymptotic symmetries and it's breaking as described in \cite{Maldacena:2016upp}. To extract the finite term, the usual procedure is to cut-off the rigid AdS near the boundary with a fixed proper length of the boundary term :
$g_{\partial AdS}= \frac{1}{\epsilon^{2}}$. The physical boundary time is parametrized by $s$ such that $t=f(s)$ and the fixed proper length boundary condition gives $u=\epsilon f'(s)$. However, the solution of the bulk Klein-Gordon equation has both normalizable and non-normalizable modes at the boundary. While the normalizable mode belongs to the solution of physical bulk fields $\Phi$, the non-normalizable mode should be thought of as a source of the boundary term $\Phi_{b}$. In more detail, the free massless scalar field mode solution in AdS$_{2}$ Poincare is $\Phi(u,t)=e^{i\omega t}\Phi_{\omega}(u)$, where $\Phi_{\omega} = c_{1}\cos[\omega z]+c_{2}\sin[\omega z]$. Hence, at $z\sim 0$, the normalizable solution of $\Phi_{\omega}$  is proportional to $\sin[\omega z]$, whereas the non-normalizable mode is just a constant. Thus without any loss of generality, we can use the following constant boundary condition at the cut-off surface
\begin{align}
    \Phi_{u=\epsilon f'(s)} =\Phi_{b}=\Phi_{\text{non-normalizable}}= \frac{\Phi_{r}}{\epsilon}
\end{align}
This particular large value of $\Phi_{b}$ is chosen to extract the finite boundary term as in \cite{Maldacena:2016upp} \footnote{Note that, in the standard JT-dilaton gravity, the solution of the equation of motion incorporates a divergent behavior of the dilaton at the asymptotic boundary. In such a case, the dilaton has no normalizable bulk solution.}. Finally using these two boundary conditions, one can extract the boundary action in terms of Schwarzian mode \cite{Mertens:2022irh}:
\begin{align}
  \frac{1}{8\pi G_{2}}\int_{\partial AdS_{2}} \sqrt{-\bar{\gamma}}\Phi_{b} (\bar{K}-1) =  \frac{\Phi_{r}}{8\pi G_{2}}\int ds \;Sch\{f(s),s\} 
\end{align}
Since we paramterize the boundary action in terms of boundary or UV time $s$, correspondingly the bulk action should also be modified in the proper boundary time frame. We may consider the following bulk coordinate transformation such that
\begin{align}
    z_{\pm} = f(\omega_{\pm}), \; \omega_{\pm} = s \pm \theta \; \text{such that at} \; \theta=\epsilon,\; t=f(s), u=\epsilon f'(s).
\end{align}
Correspondingly the metric transforms as:
\begin{align}
    ds^{2} = -4L_{2}^{2}\frac{ dz_{+} dz_{-}}{(z_{+}-z_{-})^{2}} \rightarrow -4L_{2}^{2} \left(\frac{\partial z_{+}}{\partial \omega_{+}}\right) \left(\frac{\partial z_{-}}{\partial \omega_{-}}\right)\frac{d\omega_{+}d\omega_{-}}{(z_{+}(\omega_{+})-z_{-}(\omega_{-}))^{2}}
\end{align}
Since the classical free massless scalar is Weyl invariant, the bulk action is also invariant in the new frame. Hence the total action becomes:
\begin{align}\label{jt coupled to matter}
   I = \frac{1}{16\pi G_{2}}\int   d^{2}\omega\sqrt{-g_{\{\omega_{+},\omega_{-}\}}} \frac{(\nabla_{\{\omega_{+},\omega_{-}\}}\Phi)^{2}}{2}  + \frac{\Phi_{r}}{8\pi G_{2}}\int ds\; Sch\{f(s),s\} + I_{top}  
\end{align}
We may think of this action as Schwarzian coupled to $c=1$ conformal matter. However a crucial difference exists between this and the standard description. In the standard description of JT coupled to conformal matter, the external conformal fields does not couple to dilaton and the classical equation of motion relates the dilaton with the conformal fields. However the dilaton boundary condition reduces the equation of motion to a constraint equation connecting $\frac{d( Sch\{f(s),s\})}{ds}$ to $T_{s\theta}$ at $\theta=0$. Taking $G_{2}\rightarrow0$ with $cG_{2}<<1$, the constraint gives $Sch\{f(s),s\}$ to be a constant. Hence the on-shell action gives constant Schwarzian piece which is dominated by AdS$_{2}$ black hole solution. 

On the other hand, in the present picture of (\ref{jt coupled to matter}), the matter itself plays the role of dilaton and it's non normalizable mode provides the source of the Schwarzian action. Hence the equation of motion obtained by varying the action with respect to $f(s)$, one again find $Sch\{f(s),s\}$ to be constant. Thus the classical action reduces to the free scalar field in AdS$_{2}$ black hole background with topological part and constant Schwarzian part. In \cite{Das:2025cuq}, we have argued the deformed CFT with a specific Hamiltonian generates the UV time $s$ of one-sided AdS$_{2}$ black hole, for which the thermal entropy\footnote{Since the deformed Hamiltonian is identified as the CFT modular Hamiltonian on UHP, the same entropy measures the entanglement entropy of an interval that covers a spatial slice of one-sided AdS$_{2}$ black hole.} contributes to the desired near extremal entropy\footnote{Neglecting the extremal entropy part coming from $I_{top}$.} in the classical limit. In particular it was found that the entropy $S=\frac{\Phi_{r}}{8\pi G_{2}}Sch\{f(s).s\} = \frac{c\Lambda}{12} Sch\{f(s),s\}$, with the Schwarzian to be constant\footnote{Here $c=1$.}. Here the stretched horizon cut-off $\Lambda$ plays the role of $\frac{\Phi_{r}}{G_{2}}$ and the semiclassical limit $\Phi_{r} \rightarrow\infty$ or $G_{2}\rightarrow0$ is directed by taking $\Lambda\rightarrow \infty$. To be more precise, from the effective description of near extremal near horizon physics as free dilatonic CFT, one need to account the factor $\frac{1}{16\pi G_{2}}$ within the deformed Hamiltonian. Hence, more correctly, the spatial IR cut-off $\Lambda$ of deformed CFT plays the role of $\Phi_{r}$ upto some proportionality constant\footnote{This differs from the identification of $\Lambda$ with $\frac{1}{G_{2}}$ upto some proportionality constant as proposed in \cite{Das:2025cuq}.}. Hence the classical description of Schwarzian coupled to $c=1$ CFT as described by (\ref{jt coupled to matter}), has a direct correspondence to classical limit of the same CFT with a deformed Hamiltonian \cite{Das:2025cuq}.
To summarize we may think of this in two ways:
\begin{itemize}
    \item Either, we may think the worldsheet deformed CFT is dual to the boundary Schwarzian theory with similar density of states(with proper energy rescaling).
    \item Or, the JT coupled to free bosonic CFT with Dirichlet boundary condition on scalar fields, is described by the same worldsheet deformed CFT.
\end{itemize}
In either description, the Hilbert space of modular quantization or the Hilbert space of free scalars in stretched horizon background of AdS$_{2}$ black hole provides the UV Hilbert space. Perhaps, the 3d version of this description along with emergent EFT description outside the horizon has been realized in \cite{Burman:2023kko}, \cite{Burman:2024egy}\footnote{I thank Chethan Krishnan for emphasizing this possibility at some point while discussing \cite{Burman:2023kko}.}.

\end{document}